\documentclass[fleqn,usenatbib]{mnras}

\usepackage{newtxtext,newtxmath}

\usepackage[T1]{fontenc}

\DeclareRobustCommand{\VAN}[3]{#2}
\let\VANthebibliography\thebibliography
\def\thebibliography{\DeclareRobustCommand{\VAN}[3]{##3}\VANthebibliography}

\DeclareRobustCommand{\DE}[3]{#2}
\let\DEthebibliography\thebibliography
\def\thebibliography{\DeclareRobustCommand{\DE}[3]{##3}\DEthebibliography}

\usepackage{graphicx}	
\usepackage{amsmath}	

\usepackage{booktabs}
\usepackage{float}
\usepackage{threeparttable}
\usepackage{longtable}
\usepackage{pdflscape}
\usepackage{breqn}
\usepackage{tikz,xcolor,hyperref}

\definecolor{lime}{HTML}{A6CE39}
\DeclareRobustCommand{\orcidicon}{%
    \begin{tikzpicture}
    \draw[lime, fill=lime] (0,0) 
    circle [radius=0.16] 
    node[white] {{\fontfamily{qag}\selectfont \tiny ID}};
    \draw[white, fill=white] (-0.0625,0.095) 
    circle [radius=0.007];
    \end{tikzpicture}
    \hspace{-2mm}
}

\newcommand{\orcidmpr}{\href{https://orcid.org/0000-0001-9164-2882}{\orcidicon}}
\newcommand{\orcidnad}{\href{https://orcid.org/0000-0003-4842-8834}{\orcidicon}}

\title[Heavy elements in barium stars]{Heavy elements in barium stars}

\author[M. P. Roriz et al.]{
M. P. Roriz\orcidmpr$^{1}$\thanks{E-mail: michelle@on.br}, 
M. Lugaro$^{2,3,4}$\thanks{E-mail: maria.lugaro@csfk.org}, 
C. B. Pereira$^{1}$\thanks{E-mail: claudio@on.br}, 
C. Sneden$^{5}$, 
S. Junqueira$^{1}$, 
A. I. Karakas$^{4,6}$ 
\newauthor and N. A. Drake\orcidnad$^{7,8}$
\\
$^{1}$ Observat\'orio Nacional/MCTI, Rua Gen. Jos\'e Cristino, 77, 20921-400, Rio de Janeiro, Brazil\\
$^{2}$ Konkoly Observatory, Research Centre for Astronomy and Earth Sciences, Eötvös Loránd Research Network (ELKH), H-1121 Budapest, \\Konkoly Thege M. \'ut 15-17, Hungary\\
$^{3}$ ELTE E\"{o}tv\"{o}s Lor\'and University, Institute of Physics, Budapest 1117, P\'azm\'any P\'eter s\'et\'any 1/A, Hungary\\
$^{4}$ School of Physics and Astronomy, Monash University, VIC 3800, Australia\\
$^{5}$ Department of Astronomy and McDonald Observatory, The University of Texas, Austin, TX 78712, USA\\
$^{6}$ Centre of Excellence for Astrophysics in Three Dimensions (ASTRO 3D), Melbourne, 3000 Victoria, Australia\\
$^{7}$ Laboratory of Observational Astrophysics, Saint Petersburg State University, Universitetski pr. 28, 198504, Saint Petersburg, Russia\\
$^{8}$ Laborat\'orio Nacional de Astrof\'{\i}sica/MCTI, Rua dos Estados Unidos 154, Bairro das Na\c c\~oes, 37504-364, Itajub\'a, Brazil\\
}

\date{Accepted XXX. Received YYY; in original form ZZZ}

\pubyear{2021}

\begin{document}
\label{firstpage}
\pagerange{\pageref{firstpage}--\pageref{lastpage}}
\maketitle

\begin{abstract}
New elemental abundances for the neutron-capture elements Sr, Nb, Mo, Ru, La, Sm, and Eu are presented for a large sample of 180 barium (Ba) giant stars, a class of chemically peculiar objects that exhibit in their spectra enhancements of the elements created by the $s$-process, as a consequence of mass transfer between the components of a binary system. The content of heavy elements in these stars, in fact, points to nucleosynthesis mechanisms that took place within a former asymptotic giant branch (AGB) companion, now an invisible white dwarf. From high-resolution ($R=48000$) spectra in the optical, we derived the abundances either by equivalent width measurements or synthetic spectra computations, and compared them with available data for field giant and dwarf stars in the same range of metallicity. A re-determination of La abundances resulted in [La/Fe] ratios up to 1.2 dex lower than values previously reported in literature. The program Ba stars show overabundance of neutron-capture elements, except for Eu, for which the observational data set behave similarly to field stars. Comparison to model predictions are satisfactory for second-to-first $s$-process peak ratios (e.g., [La/Sr]) and the ratios of the predominantly $r$-process element Eu to La. However, the observed [Nb,Mo,Ru/Sr] and [Ce,Nd,Sm/La] ratios show median values higher or at the upper limits of the ranges of the model predictions. This unexplained feature calls for new neutron capture models to be investigated.
\end{abstract}




\section{Introduction}\label{sec:intro}

The landmark nucleosynthesis paper by \citet{b2fh} provided us with a fundamental understanding of the different mechanisms operating within stars and responsible for creating the wide variety of chemical elements that make up the periodic table. Since then, many  studies in the literature have contributed significantly to the current comprehension of stellar nucleosynthesis \citep{wallerstein1997}. Elements beyond the iron-group ($Z>30$) are synthesized mostly by neutron captures, which are traditionally classified in two different regimes: the \textit{rapid} process \citep[$r$-process;][]{cowan2021} and the \textit{slow} process \citep[$s$-process;][]{kappeler2011}, depending on the time scale between neutron capture and $\beta$ decay of unstable nuclei on the neutron-capture path. These two mechanisms take place in different astrophysical environments, and are responsible for producing the cosmic abundance of the chemical elements between Fe and Pb in the Universe. The $r$-process occurs under explosive conditions, such as neutron star mergers, whereas the $s$-process occurs under hydrostatic conditions, in the deep layers of massive stars and low-mass stars during their asymptotic giant branch (AGB) phase \citep{herwig2005,karakas2014}.

To investigate the nucleosynthesis of the elements heavier than iron, chemically peculiar stars are essential pieces of evidence, since they record in their atmospheres the nucleosynthesis conditions of the neutron captures. Among such objects, an interesting class are the Barium (Ba) stars, first identified by \citet{bk1951}. Ba stars are G/K spectral-type giants or dwarfs with effective temperature between 4000 and 6000 K, exhibiting in their spectra strong atomic absorption lines of the elements created by the $s$-process, in particular, Ba\,{\sc ii} and Sr\,{\sc ii}, as well as molecular band features of CH, CN and C$_{2}$. The overabundance of heavy elements in a Ba star is not due to self-enrichment, because the star has not yet evolved enough to internally synthesize these elements. Such enrichment is instead a consequence of the mass transfer mechanism between the components of a binary system, in which the more evolved primary star transfers, via stellar winds, the $s$-rich material previously processed in the AGB phase to the less evolved secondary star, observed today as a Ba star. The former AGB star is now an undetectable white dwarf. Indeed, all Ba stars belong to binary systems \citep[][]{mcclure1980, mcclure1983, mcclure1990} and, thanks to the Gaia survey \citep[][]{gaiasurvey}, providing accurate measurements of radial velocity, this scenario has been confirmed and extended \citep[][]{jorissen2019, escorza2019}.

AGB stars evolve from low to intermediate mass ($1-8$ M$_{\odot}$) and their structures are composed of inert C/O cores surrounded by 
extensive H-rich convective envelopes. They are characterized by the burning of H and He alternately in two deep layers located on top of the C/O core and separated by a thin He-rich inter-shell region, where the $s$-process takes place. H-shell burning remains active most of the time ($\sim 10^{4}$ years), during the so-called interpulse period, being interrupted by brief and recurrent He-shell burning events, known as thermal pulses (TPs). After each TP, the newly synthesised $s$-process material (and also C) is dredged to the stellar surface by a mixing mechanism called third dredge-up (TDU). Two possible neutron sources drive the $s$-process in AGB stars: the $^{13}$C($\alpha$,n)$^{16}$O and $^{22}$Ne($\alpha$,n)$^{25}$Mg reactions \citep{straniero1997, gallino1998, goriley2000, busso2001, lugaro2003, karakas2014}. The $^{13}$C($\alpha$,n)$^{16}$O reaction is efficiently activated in low-mass ($\lesssim 3$ M$_{\odot}$) AGB stars during the interpulse periods, within a thin region of the intershell (the $^{13}$C pocket) at temperatures $T \sim 10^{8}$ K, providing a neutron density $N_{n} \sim 10^{7}$ cm$^{-3}$ under radiative conditions. The $^{22}$Ne($\alpha$,n)$^{25}$Mg reaction instead is only marginally activated during the TP for this mass range. For intermediate-mass ($4-8$ M$_{\odot}$) AGB stars, the $^{22}$Ne($\alpha$,n)$^{25}$Mg reaction is activated during the TPs, at higher temperatures, $T \gtrsim 3 \times 10^{8}$ K, under convective conditions, providing neutron densities up to $N_{n} \sim 10^{12}$ cm$^{-3}$. Theoretical abundance predictions of the $s$-process in AGB stars for different metalicities, [Fe/H]\footnote{Throughout this work, we have adopted the standard spectroscopy notation, [A/B] $=\log(n_{\textrm{A}}/n_{\textrm{B}}) -\log(n_{\textrm{A}}/n_{\textrm{B}})_{\odot}$ and $\log \epsilon (\textrm{A})=\log(n_{\textrm{A}}/n_{\textrm{H}})+12$, for two generic elements, A and B, where $n$ denotes the elemental abundance by number, and the $\odot$ symbol refers to the solar values.}, and masses are available, for example, from the {\sc fruity}\footnote{FUll-Network Repository of Updated Isotopic Tables \& Yields; available online at \url{http://fruity.oa-teramo.inaf.it/}} models \citep{cristallo2009, cristallo2011, cristallo2015}, the Monash group \citep{fishlock2014, karakas2016, karakas2018}, the NuGrid collaboration \citep{pignatari2016, battino2016, battino2019}, and the new {\sc snuppat} models (Yag\"ue L\'opez et al., in preparation).

Observational data are necessary to test and set constraints on these $s$-process models. For this purpose, Ba stars offer us the opportunity to link theoretical predictions to observations, with the advantage of being warmer objects than AGB stars, exhibiting spectra that are easier to analyse. Many studies have provided elemental abundances for Ba stars \citep[e.g., recent papers by ][]{allen2006a, allen2006b, allen2007, pereira2009, pereira2011, katime2013, yang2016, decastro2016, karinkuzhi2018a, karinkuzhi2018b, shejeelammal2020, roriz2021} and also contributed with deep insights into the nature of these objects, investigating their binary and kinematic properties \citep[e.g.,][]{jorissen1998, jorissen2019, escorza2017, escorza2019, escorza2020}.

In particular, \citet{decastro2016} considered a large sample consisting of 182 objects, between Ba stars and candidates, and provided abundances for Na, Al, $\alpha$-elements, iron-peak elements, and the $s$-elements Y, Zr, La, Ce, and Nd. The criterion to classify an object as a Ba star is not universally defined; \citeauthor{decastro2016} adopted the condition [$s$/Fe] $\ge$ 0.25 dex, where [$s$/Fe] is the average of the $s$-elements abundance. From that data set, \citet{cseh2018} combined among \textit{heavy} (hs) and \textit{light} (ls) $s$-elements, belonging respectively to the second (Ce and Nd) and first (Y and Zr) $s$-process peaks, to compare their ratios with predictions of the $s$-process models. \citeauthor{cseh2018} found a good agreement between the trend of observational data and models (see their Fig. 6), which confirms that $^{13}$C operates as the main neutron source in low-mass AGB stars and that metallicity is the main factor shaping the observed abundance pattern. Additionally, \citet{roriz2021} provided Rb abundances, a key element for neutron density diagnostic of the $s$-process, for the same sample of Ba stars and compared the [Rb/Zr] ratios with theoretical predictions of the $s$-process (see their Fig. 4). The observations showed a deficient Rb content (i.e., [Rb/Zr] $<$ 0) in the Ba stars, pointing to a low-neutron density for the $s$-process and consequently the low-mass nature for the former AGB companion stars, where $^{13}$C neutron source does not efficiently activate the branching points at $^{85}$Kr and $^{86}$Rb along the $s$-process path. \citet{karinkuzhi2018b} also provided Rb abundances for 10 Ba stars and \citet{shejeelammal2020} for 4 Ba stars; both studies also found [Rb/Zr] $<$ 0 for Ba stars.

Abundances for more chemical species are still needed for the understanding of nucleosynthesis in AGB stars. For this reason, in the present work, we report new elemental abundances for the neutron-capture species Sr, Nb, Mo, Ru, La, Sm, and Eu in a large sample of Ba stars, continuing the previous studies of \citet{decastro2016} and \citet{roriz2021} in Ba stars. We compare the observational data with recent theoretical predictions of the $s$-process, as well as with abundances for field stars available in literature. In the following sections, we present the sample of objects considered in this study (Section \ref{sec:sample}) and the methodology adopted to derive the chemical abundances in the program stars (Section \ref{sec:methods}); we discuss the results obtained (Section \ref{sec:discussion}) and compare the abundance ratios with prediction of the $s$-process models (Section \ref{sec:comparison}); in the end, we highlight our conclusions (Section \ref{sec:conclusions}).

\section{Target stars}\label{sec:sample}

The sample of stars considered in this study is the same as previously analyzed by \citet{roriz2021}; it is composed of the Ba stars from \citet{decastro2016}, along with 11 Ba stars from \citet{pereira2011}, 2 Ba stars from \citet{katime2013}, and 1 Ba star from \citet{pereira2009}. The target stars were observed between the years 1999 and 2010, having their high resolution spectra in the optical region obtained with the Fiber-fed Extended Range Optical Spectrograph \citep[FEROS;][]{kaufer1999} at the 1.52 m and 2.2 m ESO telescopes at La Silla (Chile). FEROS has a resolving power $R=\lambda/\Delta\lambda=48000$, covering the spectral region between 3800 \AA\ and 9200 \AA.

\section{Methods}\label{sec:methods}

\subsection{Line list selection}
 
Elemental abundances for Sr, Nb, Mo, Ru, La, Sm, and Eu were determined from a selection of atomic absorption lines of these species that are available in the spectra of the program stars. Table \ref{tab:linelist} lists the atomic data involved in the transitions, as wavelength, $\log$ \textit{gf}, and excitation potential ($\chi$). The data were taken from different sources of the literature and also from the database of the National Institute of Standards and Technology\footnote{Available online at \url{ https://physics.nist.gov/asd}} \citep[{\sc nist};][]{kramida2020} and Vienna Atomic Line Database\footnote{Available online at \url{http://vald.astro.uu.se/}} \citep[{\sc vald};][]{piskunov1995, ryabchikova2015}, as identified in the last column of the Table \ref{tab:linelist}. We have chosen atomic lines sufficiently unblended to yield reliable abundances; lines of other heavier elements require synthetic spectrum analysis with careful consideration of the strengths of contaminant transitions. In general it is best to use transition probabilities from just one atomic physics source, but that was only possible for Sm\,{\sc ii} in our work. In particular, we note that for the two La\,{\sc ii} transitions at 6320.43 \AA\ and 6774.33 \AA, \citet{lawler2001a} did not report \textit{gf}-values, so that we have adopted \textit{gf}-values from other sources, as listed in Table \ref{tab:linelist}.

\begin{table}
\centering
\caption{Line lists and atomic data considered in this study.}
\label{tab:linelist}
    \begin{threeparttable}
        \begin{tabular}{ccccc}
        \toprule
        Element & $\lambda$ (\AA) & $\chi$ (eV) & $\log$ \textit{gf}  &  Ref. \\
        \midrule
        Sr\,{\sc i}  & 4607.34  & 0.00  & $+$0.28  & S96          \\
        Sr\,{\sc i}  & 4872.49  & 1.80  & $-$0.07  & {\sc nist}   \\
        Sr\,{\sc i}  & 7070.07  & 1.85  & $-$0.03  & K18          \\
        \midrule
        Nb\,{\sc i}  & 4606.756 & 0.348 & $-$0.370 & {\sc vald}  \\
        Nb\,{\sc i}  & 5344.158 & 0.348 & $-$0.730 & {\sc vald}  \\
        Nb\,{\sc i}  & 5350.722 & 0.267 & $-$0.910 & {\sc vald}  \\
        \midrule
        Mo\,{\sc i}  & 5506.49  & 1.33  & $+$0.060 & V15    \\
        Mo\,{\sc i}  & 5533.03  & 1.33  & $-$0.069 & V15    \\
        Mo\,{\sc i}  & 5570.44  & 1.33  & $-$0.337 & V15    \\
        Mo\,{\sc i}  & 5632.46  & 1.36  & $-$1.314 & {\sc vald}\\ 
        Mo\,{\sc i}  & 5751.41  & 1.42  & $-$1.014 & {\sc vald}\\   
        Mo\,{\sc i}  & 5791.84  & 1.42  & $-$1.046 & {\sc vald}\\   
        Mo\,{\sc i}  & 5858.27  & 1.47  & $-$0.995 & {\sc vald}\\   
        Mo\,{\sc i}  & 6030.63  & 1.53  & $-$0.445 & V15    \\ 
         \midrule
        Ru\,{\sc i}  & 4757.86  & 0.93  & $-$0.539 &  A07   \\
        Ru\,{\sc i}  & 4869.15  & 0.93  & $-$0.830 &  {\sc vald}\\ 
        Ru\,{\sc i}  & 5309.27  & 0.93  & $-$1.390 &  {\sc vald}\\
        Ru\,{\sc i}  & 5636.24  & 1.06  & $-$1.070 &  {\sc vald}\\
         \midrule
        La\,{\sc ii} & 5303.53  & 0.321 & $-$1.35 &  L01a   \\
        La\,{\sc ii} & 5805.77  & 0.126 & $-$1.56 &  L01a   \\
        La\,{\sc ii} & 6262.29  & 0.403 & $-$1.22 &  L01a   \\
        La\,{\sc ii} & 6320.43  & 0.173 & $-$1.52 &  S96    \\
        La\,{\sc ii} & 6774.33  & 0.126 & $-$1.71 &  VWR00  \\
        \midrule
        Sm\,{\sc ii} & 4256.394 & 0.378 & $-$0.150 & L06    \\
        Sm\,{\sc ii} & 4318.936 & 0.277 & $-$0.250 & L06    \\
        Sm\,{\sc ii} & 4329.019 & 0.184 & $-$0.510 & L06    \\
        Sm\,{\sc ii} & 4334.150 & 0.280 & $-$0.500 & L06    \\
        Sm\,{\sc ii} & 4360.713 & 0.248 & $-$0.870 & L06    \\
        Sm\,{\sc ii} & 4362.023 & 0.484 & $-$0.470 & L06    \\
        Sm\,{\sc ii} & 4420.528 & 0.333 & $-$0.430 & L06    \\
        Sm\,{\sc ii} & 4421.133 & 0.378 & $-$0.489 & L06    \\
        Sm\,{\sc ii} & 4424.321 & 0.484 & $+$0.140 & L06    \\
        Sm\,{\sc ii} & 4433.887 & 0.434 & $-$0.190 & L06    \\
        Sm\,{\sc ii} & 4452.722 & 0.277 & $-$0.410 & L06    \\
        Sm\,{\sc ii} & 4467.341 & 0.659 & $+$0.150 & L06    \\
        Sm\,{\sc ii} & 4499.475 & 0.248 & $-$0.870 & L06    \\
        Sm\,{\sc ii} & 4523.909 & 0.434 & $-$0.390 & L06    \\
        Sm\,{\sc ii} & 4566.202 & 0.333 & $-$0.590 & L06    \\
        Sm\,{\sc ii} & 4577.690 & 0.248 & $-$0.650 & L06    \\
        Sm\,{\sc ii} & 4676.900 & 0.040 & $-$0.870 & L06    \\
        Sm\,{\sc ii} & 4704.400 & 0.000 & $-$0.860 & L06    \\
        Sm\,{\sc ii} & 4791.600 & 0.100 & $-$1.440 & L06    \\
        Sm\,{\sc ii} & 4972.170 & 0.933 & $-$0.940 & L06    \\
        \midrule
        Eu\,{\sc ii} & 6645.10  & 1.379 & $+$0.12  & L01b   \\
\bottomrule               
\end{tabular}
        \textbf{References:}
        S96: \citet{sneden1996}; K18: \citet{karinkuzhi2018b}; V15: \citet{veklich2015}; A07: \citet{allen2007}; L01a: \citet{lawler2001a}; VWR00: \citet{winckel00}; L06: \citet{lawler2006}; L01b: \citet{lawler2001b}.
    \end{threeparttable}
\end{table}

In this study we have taken into account hyperfine splitting (HFS) components to derive the abundances of Nb, La, and Eu. For Eu\,{\sc ii} line at 6645.10 \AA, we considered HFS data of \citet{lawler2001b}; for La\,{\sc ii} lines at 5303.53 \AA, 5805.77 \AA, and 6262.29 \AA, HFS data are from \citet{lawler2001a}. For the lines at 6320.43 \AA\ and 6774.33 \AA\ of La\,{\sc ii}, as well as for the Nb lines considered here, we have computed the $\log$ \textit{gf} value for each hyperfine component, following the prescriptions described in Section 4 of \citet{roriz2021}, i.e., by distributing the total $\log$ \textit{gf} of the line (given in Table \ref{tab:linelist}) according to the relative intensity of the hyperfine transitions. The HFS components and their respective $\log$ \textit{gf} are listed in Tables \ref{tab:app1} and \ref{tab:app2}.

For the program stars, we have adopted their atmospheric paramenters as recommended by \citet{decastro2016}, \citet{pereira2011}, and \citet{katime2013}. The atmospheric parameters reported by these studies were all obtained from the local thermodynamic equilibrium (LTE) plane-parallel atmosphere models provided by \citet{kurucz1993}, and the LTE code of spectral analysis {\sc moog} \citep{sneden1973}, adopting the usual technique, in which the effective temperature ($T_{\textrm{eff}}$) of the star is fixed from the excitation equilibrium of the Fe\,{\sc i} lines, whereas the ionisation equilibrium between Fe\,{\sc i} and Fe\,{\sc ii} ions constraints the superficial gravity ($\log g$); the independence beteween the Fe\,{\sc i} abundances and the reduced equivalent widths ($W_{\lambda}/\lambda$) of their spectral lines provides the microturbulent velocity ($\xi$).

\subsection{Abundance determination}

Abundances for the atomic species considered in this study were derived either by comparing observed and synthetic spectra or by equivalent width measurements of the selected lines. To compute the abundances, we have used the current version of the code {\sc moog}\footnote{Available at \url{https://www.as.utexas.edu/~chris/moog.html}}, running the drivers \textit{abfind}, \textit{synth}, and \textit{blends}. Throughout this study, we have adopted the photosphere solar abundances as recommended by \citet{grevesse1998}. The new elemental abundances of the chemical species considered in this work are listed in Table \ref{tab:app3} and will be discussed in Section \ref{sec:discussion}.

\begin{figure*}
    \centering
        \includegraphics{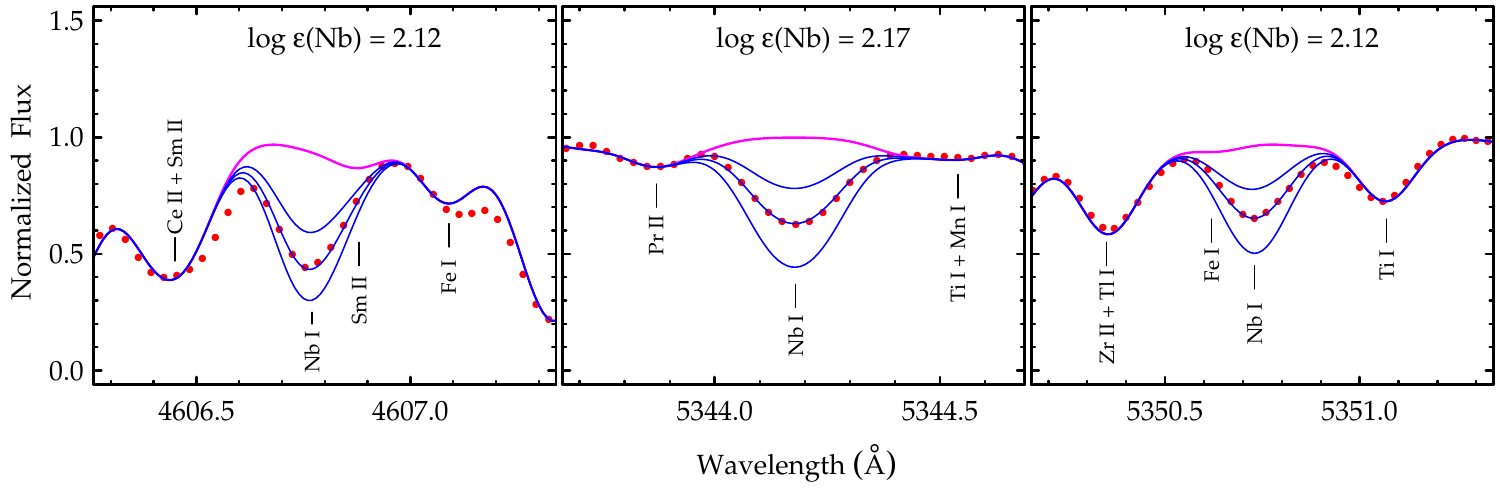}
        \caption{Small spectral regions surrounding Nb\,{\sc i} lines at 4606.77 \AA, 5344.16 \AA, and 5350.72 \AA\ for the star HD 67036, showing the observed (red dots) and synthetic spectra (colored curves). In each panel, magenta curve is the synthetic spectrum with no Nb contribution; the best fit between observed and computed spectra gives the Nb abundance, labeled on top of the panels. Synthetic spectra for $\Delta \log \epsilon (\textrm{Nb})= \pm 0.3$ dex around the best fit are also shown in each panel. The adopted abundance is obtained by averaging the values coming from the best fits.}
        \label{fig:synnb}
\end{figure*}

For the elements Sr, Mo, Ru, and Sm, we measured the equivalent widths of the atomic lines in Table \ref{tab:linelist} by fitting a Gaussian profile in the observed spectra and then, from the driver \textit{abfind}, we were able to determine their elemental abundances, since these lines are free of any contamination. For the elements Nb and Eu, on the other hand, our analysis was based on spectral synthesis technique, from the driver \textit{synth} of {\sc moog}, taking into account the HFS splits, as previously mentioned. We show in Fig. \ref{fig:synnb} an example of spectral synthesis for the Nb lines considered in the present study, as well as we indicate other atomic contributions close to the region of the lines.

Although La abundances for the program stars were already reported by the previous studies of \citet{decastro2016}, \citet{pereira2011}, \citet{katime2013}, and \citet{pereira2009}, from which we selected our sample, the analyses in these papers did not consider the HFS splits for La\,{\sc ii} lines. Such single-line approaches are inadequate for these transitions, which are strongly affected by HFS interaction, and result in overestimated [La/Fe] ratios, when compared with the ratios of their neighborhood elements, as [Ce/Fe] and [Nd/Fe]. Since they all belong to the same $s$-process peak (i.e., second peak), these species are expected to have similar overabundances and $s$-process models are not able to reach the high [La/Ce] ratios. For this reason, La was not considered in the work of \citet{cseh2018}. In the present study, we have re-computed the La abundances with full HFS corrections. From equivalent width measurements of the La\,{\sc ii} atomic lines considered, we run the driver \textit{blends} of {\sc moog}, which provides abundances from blended spectral lines.

\begin{figure*}
    \centering
        \includegraphics{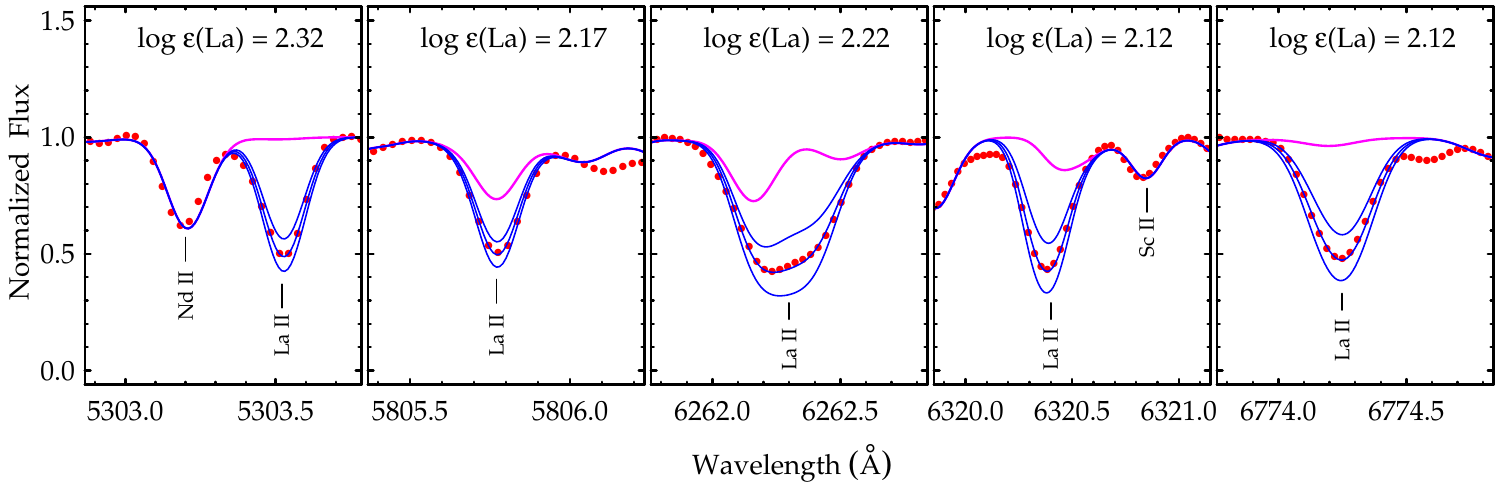}
        \caption{Small spectral regions surrounding La\,{\sc ii} lines at 5305.53 \AA, 5805.77 \AA, 6262.29 \AA, 6320.43 \AA, and 6774.33 \AA\ for the star HD 24035, showing the observed (red dots) and synthetic spectra (colored curves). In each panel, magenta curve is the synthetic spectrum with no La contribution; the best fit between observed and computed spectra gives the La abundance, labeled on top of the panels. Synthetic spectra for $\Delta \log \epsilon (\textrm{La})= \pm 0.3$ dex around the best fit are also shown in each panel. The adopted abundance is obtained by averaging the values coming from the best fits.}
        \label{fig:synla}
\end{figure*}

The revised values for the [La/Fe] ratios presented here are up to $\sim$ 1.2 dex lower than those previously reported by \citet{decastro2016}, consistent with the [Ce/Fe] and [Nd/Fe] ratios, which allow us to compare them with theoretical predictions of the $s$-process, as we will discuss in Section \ref{sec:comparison}. For instance, the highest La abundance reported by \citeauthor{decastro2016} is [La/Fe] = 2.70 dex, for star HD 24035; in this study, however, we have derived [La/Fe] = 1.45 $\pm$ 0.21 dex for this object. Only for comparison purposes, we selected from the data set of \citeauthor{decastro2016} all the objects displaying [La/Fe] $>$ 2.0 dex and then we computed synthetic spectra for the stars belonging to this sub sample. We show in Fig. \ref{fig:synla} the observed and synthetic spectra computed for HD 24035, giving [La/Fe] = 1.25 $\pm$ 0.21 dex, in good agreement with the adopted value in this study, presented in Table \ref{tab:app3}, obtained from the measurements of the equivalent width. We can also see in Fig. \ref{fig:synla}, from the synthetic spectra without La contribution, that the possible contaminants for the La lines are not significant to its final abundance.

\subsection{Abundances uncertainties}

The uncertainties in abundances come from errors in the parameters of the atmospheric models ($T_{\textrm{eff}}$, $\log g$, $\xi$, and [Fe/H]) and from errors in equivalent width ($W_{\lambda}$) measurements and synthetic/observed spectrum matches; in addition, the dispersion in abundance due to the number of lines considered in its derivation is also another error source.  Combining quadratically each one of these terms, we can estimate the total uncertainty in abundances according to the following equation:

\begin{dmath}\label{eq:eq3}
     \sigma_{\log \epsilon(\textrm{X})_{\star}}^{2}=\sigma_{\textrm{ran}}^{2}+\left(\frac{\partial \log \epsilon}{\partial T_{\textrm{eff}}}\right)^{2}\sigma_{T_{\textrm{eff}}}^{2}+ \left(\frac{\partial \log \epsilon}{\partial \log g}\right)^{2}\sigma_{\log g}^{2}+ \left(\frac{\partial \log \epsilon}{\partial \xi}\right)^{2}\sigma_{\xi}^{2}+ \left(\frac{\partial \log \epsilon}{\partial \textrm{[Fe/H]}}\right)^{2}\sigma_{\textrm{[Fe/H]}}^{2}+ \left(\frac{\partial \log \epsilon}{\partial W_{\lambda}}\right)^{2}\sigma_{W_{\lambda}}^{2}, 
\end{dmath}

\noindent where the partial derivatives correspond to variations in abundance when we change one parameter, keeping the others constant, and $\sigma_{\textrm{ran}}=\sigma_{\textrm{obs}}/\sqrt{N}$, where $\sigma_{\textrm{obs}}$ is the standard deviation and $N$ is the number of lines considered to derive the abundance of a given chemical species X. The uncertainty in the [X/Fe] ratio is given by:

\begin{equation}
    \sigma_{\textrm{[X/Fe]}}^{2}=\sigma_{\textrm{X}}^{2}+\sigma_{\textrm{Fe}}^{2}.
\end{equation}

Following the same approach as adopted by \citet{cseh2018} and \citet{roriz2021}, we group the program stars into three temperature ranges: Group 1 (5000 - 5400 K), Group 2 (4700 - 4950 K), and Group 3 (4100 - 4600 K), based respectively on Tables 9, 10, and 11 of \citet{decastro2016}, where each group is represented by a typical star, for which we have evaluated the partial derivatives in equation \ref{eq:eq3}. Tables \ref{tab:tab5200}, \ref{tab:tab4800}, and \ref{tab:tab4400} present the variation in abundances, when we change each one of the stellar parameters, as well as the equivalent widths. The square root of the sum of the squares of these quantities, as listed in column 7 of Tables \ref{tab:tab5200}, \ref{tab:tab4800}, and \ref{tab:tab4400}, then provides an estimate for the uncertainties in the abundances for the three temperature ranges. The quantity $\sigma_{\textrm{ran}}$, on the other hand, was computed for each star of the sample, when three or more lines were available.

We note, however, that for the representative star of Group 2, HD 119185, the three Nb lines were absent in its spectrum. Therefore, in order to estimate the uncertainties in the Nb abundance for the temperature range of 4700 - 4950 K, we selected a similar star, HD 116869, with the same effective temperature, for which two Nb lines, 5354.16 \AA\ and 5350.72 \AA\, are available. In this way, we computed a variation in the logarithmic abundance of $+0.19$ dex, for $\Delta T_{\textrm{eff}}=+100$K; $+0.02$ dex, for $\Delta \log g=+0.2$ dex; $+0.02$ dex, for $\Delta \xi=+0.3$ km/s; and $+0.02$, for $\Delta \textrm{[Fe/H]}=+0.1$ dex, giving a compounded uncertainty $\sqrt{\Sigma \sigma^{2}}=0.19$ dex. 

\begin{table*}
	\centering
	\caption{Abundance uncertainties for the star BD$-$14$^{\circ}$2678, which has $T_{\textrm{eff}}=5200$ K, $\log g=3.1$, $\textrm{[Fe/H]}=0.01$ dex, and $\xi=1.4$ km s$^{-1}$. From the second to sixth column, we show the changes in abundances due to variation in $T_{\textrm{eff}}$, $\log g$, $\xi$, [Fe/H], and $W_{\lambda}$, respectively. The seventh column gives the compounded rms uncertainties of the second to sixth column. The last column provides the abundances dispersion observed among the lines, when three or more lines are available.}
	\label{tab:tab5200}
	\begin{tabular}{lccccccc} 
		\toprule
		Species & $\Delta T_{\textrm{eff}}$ & $\Delta \log{g}$ & $\Delta \xi$ & $\Delta$[Fe/H] & $\Delta W_{\lambda}$ & $\sqrt{\Sigma \sigma^{2}}$ & $\sigma_{\textrm{obs}}$ \\
		        &   $+100$ K                &  $+0.2$          & $+0.3$ km s$^{-1}$ & $+0.1$ dex    & $+3$ m\AA & & \\           
		\midrule
        Sr\,{\sc i}  & $+$0.11 & $-$0.02 & $-$0.15 & $-$0.01 & $+$0.07 & 0.20 & ...  \\
        Nb\,{\sc i}  & $+$0.18 & $+$0.03 &    0.00 & $+$0.08 & ...     & 0.20 & ...  \\
        Mo\,{\sc i}  & $+$0.12 & $-$0.01 & $-$0.03 & $-$0.01 & $+$0.11 & 0.17 & 0.18 \\
        Ru\,{\sc i}  & $+$0.13 &    0.00 & $-$0.02 &    0.00 & $+$0.09 & 0.16 & ...  \\
        La\,{\sc ii} & $+$0.02 & $+$0.09 & $-$0.04 & $+$0.04 & $+$0.04 & 0.12 & 0.10 \\
        Sm\,{\sc ii} & $+$0.03 & $+$0.09 & $-$0.09 & $+$0.04 & $+$0.08 & 0.16 & 0.13 \\
        Eu\,{\sc ii} &  0.00   & $+$0.08 & $-$0.02 & $+$0.03 &   ...   & 0.09 & ...  \\
		\bottomrule
	\end{tabular}
\end{table*}

\begin{table*}
	\centering
	\caption{Same as Table \ref{tab:tab5200}, however for the star HD 119185, which has $T_{\textrm{eff}}=4800$ K, $\log g=2.0$, $\textrm{[Fe/H]}=-0.43$ dex, and $\xi=1.3$ km s$^{-1}$.}
	\label{tab:tab4800}
	\begin{tabular}{lccccccc} 
		\toprule
		Species & $\Delta T_{\textrm{eff}}$ & $\Delta \log{g}$ & $\Delta \xi$ & $\Delta$[Fe/H] & $\Delta W_{\lambda}$ & $\sqrt{\Sigma \sigma^{2}}$ & $\sigma_{\textrm{obs}}$ \\
		        &   $+100$ K                &  $+0.2$          & $+0.3$ km s$^{-1}$ & $+0.1$ dex    & $+3$ m\AA & & \\           
		\midrule
        Sr\,{\sc i}  & $+$0.14 & $-$0.02 & $-$0.16 & $-$0.01 & $+$0.05 & 0.22 & ...  \\
        Nb\,{\sc i}  & ...     & ...     &    ...  &  ...    &  ...    &  ... & ...  \\
        Mo\,{\sc i}  & $+$0.15 & $-$0.02 & $-$0.02 & $-$0.01 & $+$0.09 & 0.18 & 0.10 \\
        Ru\,{\sc i}  & $+$0.16 & $-$0.01 & $-$0.01 &    0.00 & $+$0.10 & 0.19 & ...  \\
        La\,{\sc ii} & $+$0.02 & $+$0.09 & $-$0.03 & $+$0.03 & $+$0.04 & 0.11 & 0.15 \\
        Sm\,{\sc ii} & $+$0.03 & $+$0.09 & $-$0.11 & $+$0.03 & $+$0.08 & 0.17 & 0.19 \\
        Eu\,{\sc ii} &    0.00 & $+$0.07 &    0.00 &    0.00 &   ...   & 0.07 & ...  \\
		\bottomrule
	\end{tabular}
\end{table*}

\begin{table*}
	\centering
	\caption{Same as Table \ref{tab:tab5200}, however for the star HD 130255, which has $T_{\textrm{eff}}=4400$ K, $\log g=1.5$, $\textrm{[Fe/H]}=-1.11$ dex, and $\xi=1.3$ km s$^{-1}$.}
	\label{tab:tab4400}
	\begin{tabular}{lccccccc} 
		\toprule
		Species & $\Delta T_{\textrm{eff}}$ & $\Delta \log{g}$ & $\Delta \xi$ & $\Delta$[Fe/H] & $\Delta W_{\lambda}$ & $\sqrt{\Sigma \sigma^{2}}$ & $\sigma_{\textrm{obs}}$ \\
		        &   $+90$ K                 &  $+0.2$          & $+0.3$ km s$^{-1}$ & $+0.1$ dex    & $+3$ m\AA & & \\           
		\midrule
        Sr\,{\sc i}  & $+$0.14 & $-$0.02 & $-$0.15 & $-$0.01 & $+$0.17 & 0.27 &  ... \\
        Nb\,{\sc i}  & $+$0.19 & $-$0.02 &    0.00 &    0.00 &   ...   & 0.19 &  ... \\
        Mo\,{\sc i}  & $+$0.15 & $-$0.02 & $-$0.03 &    0.00 & $+$0.11 & 0.19 & 0.08 \\
        Ru\,{\sc i}  & $+$0.16 & $-$0.01 & $-$0.01 & $+$0.01 & $+$0.19 & 0.25 & 0.07 \\
        La\,{\sc ii} & $+$0.02 & $+$0.09 & $-$0.01 & $+$0.04 & $+$0.05 & 0.11 & 0.10 \\
        Sm\,{\sc ii} & $+$0.03 & $+$0.09 & $-$0.10 & $+$0.03 & $+$0.09 & 0.17 & 0.09 \\
        Eu\,{\sc ii} &    0.00 & $+$0.09 &    0.00 & $+$0.03 &   ...   & 0.09 &  ... \\
		\bottomrule
	\end{tabular}
\end{table*}

\section{Discussion}\label{sec:discussion}

\begin{figure*}
    \centering
        \includegraphics{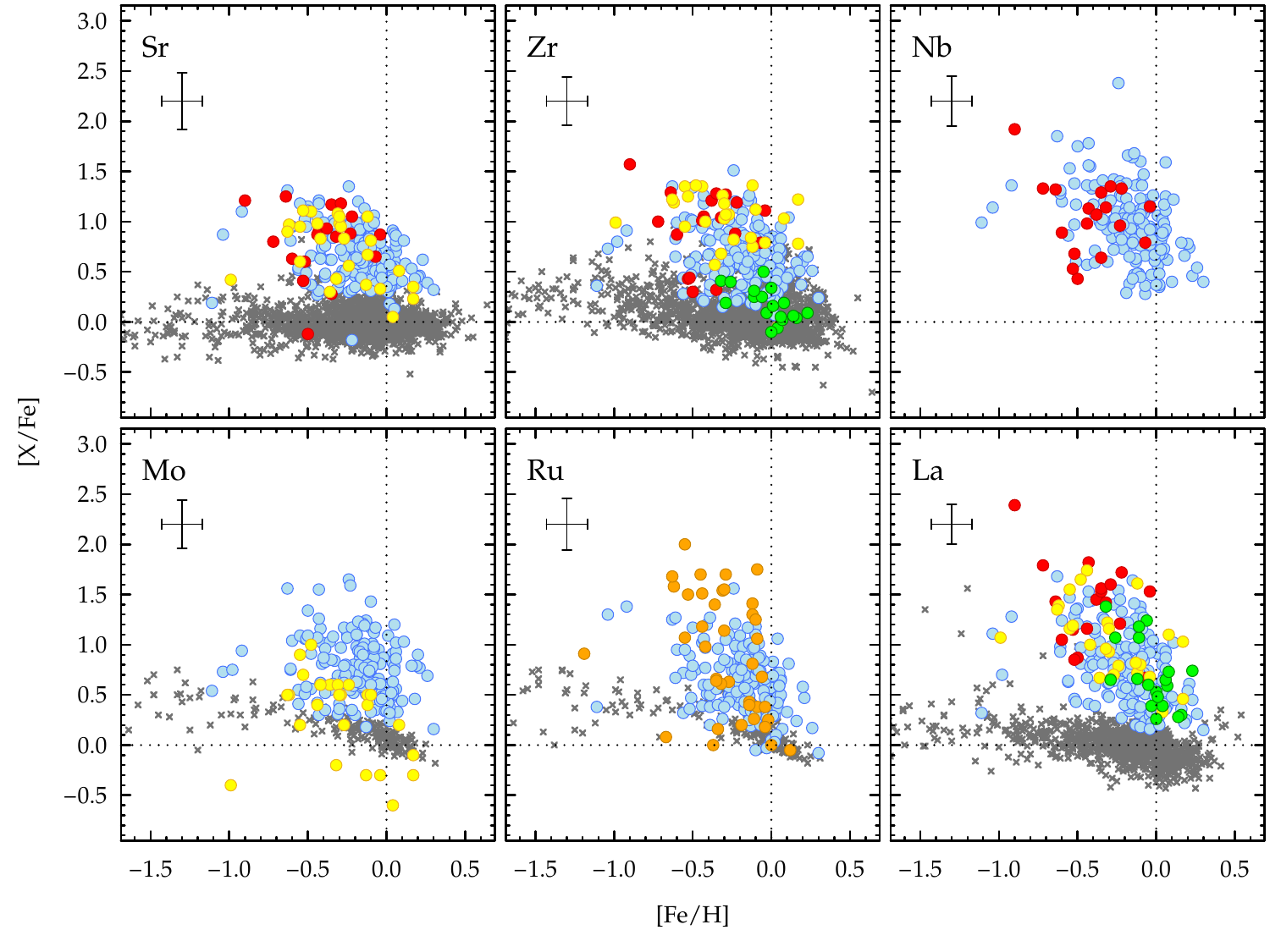}
        \caption{Observed abundance ratios, [X/Fe], against metallicity for the Ba giant stars (blue dots) considered in the present study; the Zr abundances for the program stars, also presented in this figure, were already reported by the previous studies. For sake of clarity, we have represented in the panels only typical error bars given in the Table \ref{tab:app3} for each ratio. Color dots are data for Ba stars taken from the previous works of \citet[][red]{karinkuzhi2018b}, \citet[][yellow]{allen2006a}, \citet[][orange]{allen2007}, and \citet[][green]{yang2016}. We also added in the plots abundances for field giant and dwarf stars (grey crosses), taken from different sources of literature \citep[][]{gratton1994, jehin1999, fulbright2000, mashonkina2001, mishenina2001, luck2006, luck2007, mishenina2007, mishenina2013, ishigaki2013, hansen2014, battistina2016, delgado2017, mishenina2019a, mishenina2019b, forsberg2019}.}
        \label{fig:s}
\end{figure*}

In Table \ref{tab:fraction}, we present the contribution of the $s$-process to the solar abundances of some $n$-elements, among which we have considered in this study, as reported by different works in literature. Sr, Zr, and La are the species with generally the highest contribution from the $s$-component, in contrast to Sm and Eu, which are two elements more representative of the  $r$-process. In Fig. \ref{fig:s}, we plot the ratios [X/Fe] against [Fe/H] for the elements Sr, Nb, Mo, Ru, and La, together with data for Ba stars from the previous works of \citet{allen2006a}, \citet{allen2007}, \citet{yang2016}, and \citet{karinkuzhi2018b}; we added in Fig. \ref{fig:s} the Zr abundances reported by \citet{decastro2016} for the program stars. We also compare the data sets with abundances observed in field giant and dwarf stars, taken from different available sources of the literature, as listed in the figure caption. Ba stars show a clear enrichment of $s$-elements in their atmospheres relative to the field stars; and such enrichment is prominent for Sr, Zr, and La, as expected. Nb, Mo, and Ru in the Ba stars show, in general, a similar behavior to Sr, Zr, and La, where the ratios [X/Fe] increase with decrease of metallicity. However, field stars also show this same trend for Mo and Ru, while Sr, Zr, and La abundances remain approximately constant, close to the solar values. For Nb, there is not available data in literature for field stars, and this element stands out as most of the Ba stars are above solar.

\begin{table}
\centering
\caption{Contribution (\%) of the $s$-component for the neutron-capture elements considered in this study.}
\label{tab:fraction}
    \begin{threeparttable}
        \begin{tabular}{lcccc}
        \toprule
	Species &  S96 & A99 & B00 & B14\\
	\midrule
        Sr  &   85.2 & 85  & 89  & 68.9 \\
        Y   &   71.8 & 92  & 72  & 71.9 \\
        Zr  &   83.3 & 83  & 81  & 66.3 \\
        Nb  &   67.6 & 85  & 68  & 56.0 \\
        Mo  &   67.8 & 50  & 68  & 38.7 \\
        Ru  &   39.0 & 32  & 39  & 28.9 \\
        La  &   75.2 & 62  & 75  & 75.5 \\
        Ce  &   77.5 & 77  & 81  & 83.5 \\
        Nd  &   47.1 & 56  & 47  & 57.5 \\
        Sm  &   24.2 & 29  & 34  & 31.4 \\
        Eu  &    2.7 & 5.8 & 3   &  6.0 \\
	\bottomrule
\end{tabular}
        \textbf{References:}
        S96: \citet{sneden1996}; A99: \citet{arlandini1999}; B00: \citet{burris2000}; B14: \citet{bisterzo2014}.
    \end{threeparttable}
\end{table}

\begin{figure}
    \centering
        \includegraphics{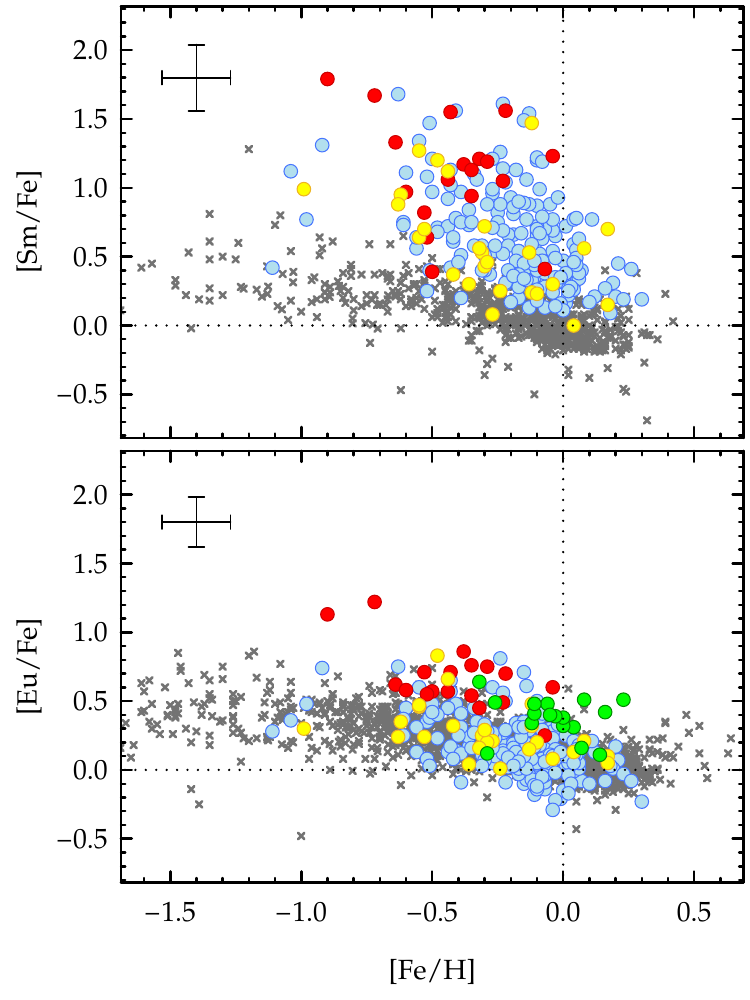}
        \caption{Abundance ratios for Sm and Eu, two representative $r$-elements, against metallicity. Symbols have the same meaning as in Fig. \ref{fig:s}. Data for field stars were taken from the same references listed in the caption of Fig. \ref{fig:s}.}
        \label{fig:sm-eu}
\end{figure}

In Fig. \ref{fig:sm-eu}, we show the [X/Fe] ratios versus [Fe/H] for the two elements mostly contributed by the $r$-process (see Table \ref{tab:fraction}), Sm and Eu. As one can see on the top panel of Fig. \ref{fig:sm-eu}, when compared with field stars, Ba stars show a degree of enrichment in Sm, displaying [Sm/Fe] ratios between $0.0$ and $+1.5$ dex, which point to the $s$-process contribution. The lower panel of Fig. \ref{fig:sm-eu} shows that Ba stars behave similar to field stars, exhibiting [Eu/Fe] between $-0.3$ and $+0.8$ dex. Considering the fact that Eu is an element with very small contribution of the $s$-process (only $3-6$\%; see Table \ref{tab:fraction}) and that the content transferred by the more evolved binary companion is diluted in the atmospheres of the Ba stars, we conclude that the Eu content in Ba stars is mainly of galactic origin. 

\begin{figure}
    \centering
        \includegraphics{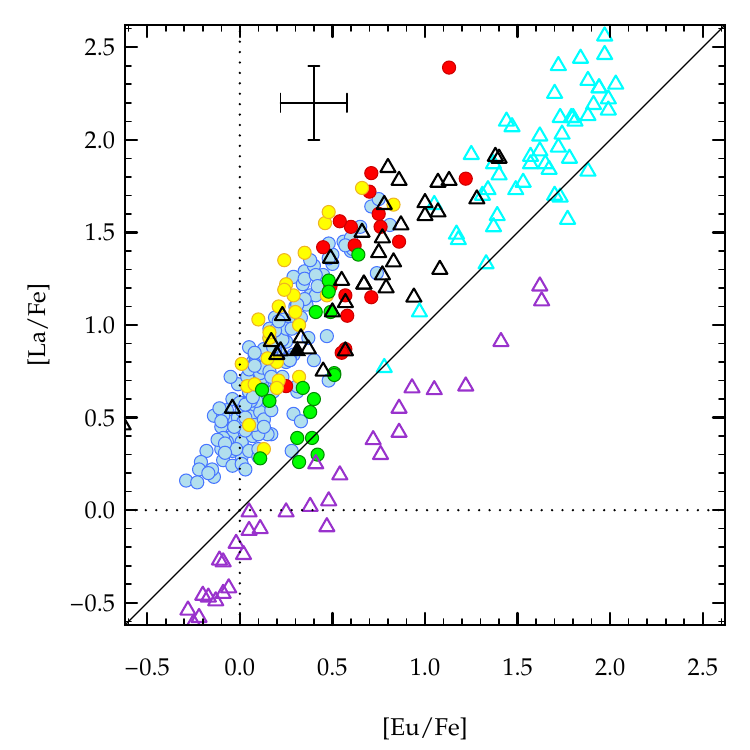}
        \caption{Ba stars and CEMP stars in the [La/Fe] versus [Eu/Fe] plane. The color dots have the same meaning as Fig. \ref{fig:s}. Triangles are data for CEMP-$s$ (black), CEMP-$r$ (purple) and CEMP-$r/s$ (cyan) stars, taken from \citet{masseron2010} and \citet{karinkuzhi2021}. The full black triangle is the CEMP-$s$ star CD$-$50$^{\circ}$776, reported by \citet{roriz2017}.}
        \label{fig:la-eu}
\end{figure}

In Fig. \ref{fig:la-eu}, we show the positions of Ba stars in the [La/Fe] versus [Eu/Fe] plane, including data of Ba stars from literature. This diagram is useful because it separates between $s$-rich and $r$-rich objects, since La and Eu are elements with strong contribution of the $s$-process ($\sim 75$\%) and $r$-process ($\sim 95$\%), respectively (Table \ref{tab:fraction}). For comparison purposes, we also add in Fig. \ref{fig:la-eu} data for another important class of chemically peculiar stars, known as Carbon-Enhanced Metal-Poor (CEMP) stars, which exhibit [C/Fe] $>1.0$ dex in the metallicity range [Fe/H] $\le -1$ dex \citep[e.g., see][and references therein]{masseron2010}. Depending on the enrichment of $s$- and $r$-elements, CEMP stars are also classified as CEMP-$s$, CEMP-$r$ or CEMP-$r/s$. Among these sub classes, the $s$-enrichment observed in CEMP-$s$ stars is attributed to mass transfer hypothesis, as in Ba stars. As one can see in Fig. \ref{fig:la-eu}, the Ba stars are located on the $s$-rich side of the plane, close to the CEMP-$s$ stars, whereas CEMP-$r$ stars are found on the $r$-rich region of the diagram, as expected. The CEMP-$r/s$ stars represent a different population, which was renamed CEMP-$i$ by \citet{hampel2016} as these stars are better matched by an intermediate neutron-capture ($i$) process.

\subsection{The Zr-Nb plane as a temperature diagnostic of the \textit{s}-process}

\citet{neyskens2015} showed that the contents of Zr and Nb observed in S stars can be used as a tool for temperature diagnostic of the $s$-process, being an indicator of the main neutron source reaction that operated inside the former AGB companion. This method was also applied by \citet{karinkuzhi2018b} to Ba stars. Assuming that (i) the amount of $^{93}$Zr, with half-live $t_{1/2} \sim 1.5$ Myr, transferred from more evolved AGB star, had time to decay into $^{93}$Nb, (ii) the Zr isotopic abundances are in local equilibrium, in which the product $\langle \sigma_{A} \rangle N_{A}$ is constant, and that (iii) the branching point at $^{95}$Zr is closed, \citeauthor{neyskens2015} showed that the expected ratios [Zr/Fe] and [Nb/Fe] should obey the following relation:

\begin{equation} \label{eq:eq4}
    \left[\dfrac{\textrm{Zr}}{\textrm{Fe}}\right]=\left[\dfrac{\textrm{Nb}}{\textrm{Fe}}\right]+\log \omega^{*}-\log \dfrac{N_{\odot}(\textrm{Zr})}{N_{\odot}(\textrm{Nb})},
\end{equation}

\noindent where the quantity $\omega^{*}$ is written in terms of the Maxwellian-averaged neutron-capture cross section (MACS), $\langle \sigma_{A}\rangle$, of the Zr isotopes: 

\begin{equation} \label{eq:eq5}
    \omega^{*}=\langle \sigma_{93} \rangle \left[ \dfrac{1}{\langle \sigma_{90} \rangle}+\dfrac{1}{\langle \sigma_{91} \rangle}+\dfrac{1}{\langle \sigma_{92} \rangle}+\dfrac{1}{\langle \sigma_{94} \rangle}\right],
\end{equation}

\noindent for which neutron-capture cross sections are available, for example, in KADoNiS\footnote{Available online at: \url{https://www.kadonis.org}} database \citep{dillmann2006}. 

Equation \ref{eq:eq4} defines a straight line of slope equal to 1 in the [Zr/Fe] versus [Nb/Fe] plane, which is shifted according to the value of $\omega^{*}$. The value of the Zr isotopes cross sections relative to each other show some temperature dependence; therefore, $\omega^{*}$ also show some temperature dependence and can point to the neutron source in operation. Considering temperatures in the interval of $(1-3)\times 10^{8}$ K, equivalent to $10-30$ keV, at which lower and higher boundary the $^{13}$C($\alpha$,n)$^{16}$O and $^{22}$Ne($\alpha$,n)$^{25}$Mg reactions act as main neutron source, respectively, then equation \ref{eq:eq4} delimits a narrow region in the [Zr/Fe] versus [Nb/Fe] plane, restricted by the values $\omega^{*}=15.4$ (for 10 keV) and $\omega^{*}=13.3$ (for 30 keV), where we  expect to find the Ba stars. Using more accurate prediction from {\sc starevol} models, for [Fe/H] = $-0.5$ dex, \citet{karinkuzhi2018b} found $\omega^{*}=15.8$ for AGB stars with masses between $2-3$ M$_{\odot}$, in good agreement with the simple analytical approach of \citet{neyskens2015}. For more massive AGB stars instead (i.e., $4-5$ M$_{\odot}$), {\sc starevol} models predict a different trend from that expected by equation \ref{eq:eq4}, as discussed in detail by \citet[][see their Section 7.4]{karinkuzhi2018b}.

\begin{figure}
    \centering
        \includegraphics{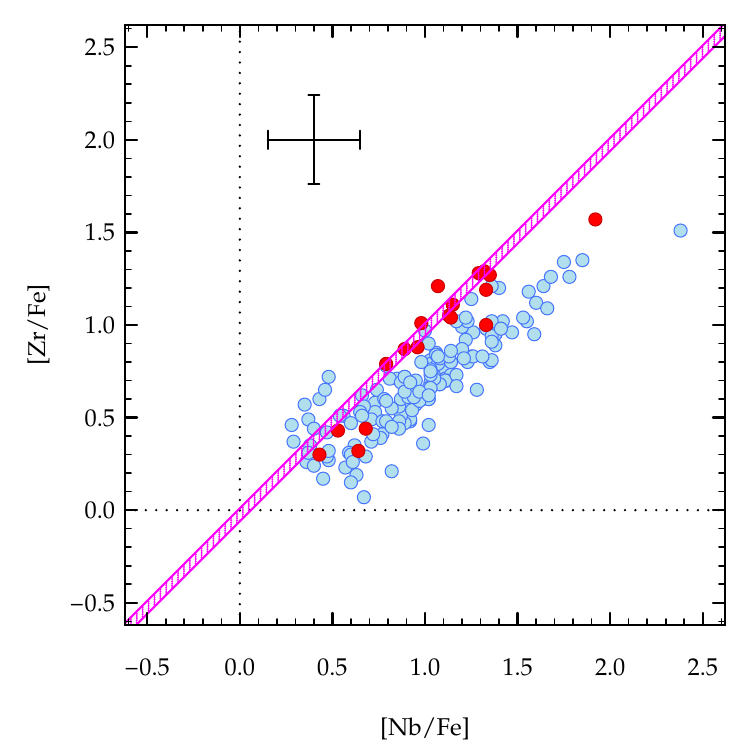}
        \caption{Ba stars in the [Zr/Fe] versus [Nb/Fe] plane. Symbols have the usual meaning, as Fig. \ref{fig:s}. The narrow magenta region is the prediction, from equation \ref{eq:eq4}, for extrinsic stars enriched by $s$-elements previously processed under temperatures between $(1-3) \times 10^{8}$ K \citep[][]{neyskens2015}.}
        \label{fig:nb-zr}
\end{figure}

In Fig. \ref{fig:nb-zr}, we plot [Zr/Fe] against [Nb/Fe], similar to Fig. 14 of \citet{karinkuzhi2018b}, where we show the magenta region delimited by the upper (for 10 keV) and lower (for 30 keV) lines. As shown in Fig. \ref{fig:nb-zr}, the target stars of this study are distributed systematically below the expected magenta region, whereas only a few points are located above that region. We added in Fig. \ref{fig:nb-zr} the sample of Ba stars considered by \citet{karinkuzhi2018b}; their data partially follow the expected relation predicted by equation \ref{eq:eq4}, although some of their points are also located below the magenta region. We conclude that our stars show higher Nb than expected by the models, and this point is discussed in more detail in Section \ref{sec:comparison}.

\subsection{Indexes of the \textit{s}-process}

\begin{figure*}
    \centering
    \includegraphics{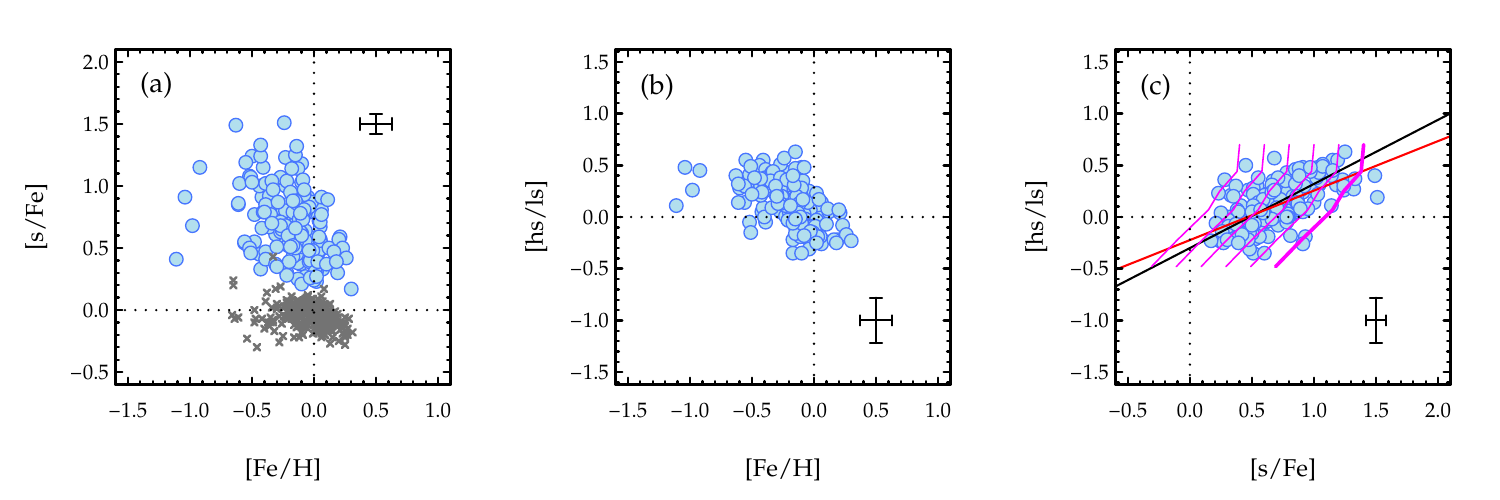}
    \caption{Indexes of the $s$-process for the Ba giant stars considered in this study. In panel (a) we compare the average of the $s$-process for Ba stars (blue dots) and field giant stars (grey crosses); in panel (b) we consider the index [hs/ls] versus [Fe/H]; in panel (c) we show the correlation between [hs/ls] and [$s$/Fe] for the program stars. The red straight line is a fit by least-squares, whereas the black straight line is the fit derived by \citet{decastro2016}. The magenta thick curve represents the predicted trend from {\sc fruity} AGB models of 2 M$_{\odot}$ and different metallicities (with higher metallicities predicting higher [hs/ls] and [$s$/Fe]; see text for discussion). The magenta thin curves are the model predictions shifted towards lower [$s$/Fe] by steps of 0.2 dex to mimic the effect of dilution on the secondary Ba stars, which affects (decreases) [$s$/Fe] but not [hs/ls].}
    \label{fig:s_mean}
\end{figure*}

In panel (a) of Fig. \ref{fig:s_mean}, we plot the average of the $s$-process for the Ba giant stars, considering the elements Rb, Sr, Y, Zr, Nb, Mo, Ru, La, Ce, and Nd, and compare the updated [$s$/Fe] ratios with data for field giant stars taken from \citet{luck2007} and \citet{mishenina2007}. As we mentioned before, the program stars are selected such as [$s$/Fe] $\ge 0.25$ dex, although we have also considered four objects of the sample, HD 49017, HD 119650, HD 212209, and MFU 214, for which we derive [$s$/Fe] = 0.23, 0.21, 0.17, and 0.24 dex, respectively, as these values are still $\ge 0.25$ dex within error bars. In panel (b), we plot the index [hs/ls] versus [Fe/H], where [hs/ls] = [hs/Fe] - [ls/Fe], where [hs/Fe] and [ls/Fe] are respectively the averages of the heavy (La, Ce and Nd) and light (Sr, Y and Zr) $s$-elements. As one can see in panel (b), the [hs/ls] ratio is anti-correlated with metallicity, as discussed previously by \citet{cseh2018}. Finally, in panel (c), we plot [hs/ls] against [$s$/Fe] and fitted to data a straight (in red) by least-squares, giving [hs/ls] $=(-0.22 \pm 0.03)+(0.48 \pm 0.05) \times$ [$s$/Fe].

In Fig. \ref{fig:s_mean} panel (c) we also show the AGB predicted trend using as example {\sc fruity} AGB models of 2 M$_{\odot}$ and different metallicities. The main driver of the predicted [hs/ls] and [$s$/Fe] correlation is the metallicity of the star. In fact, lower metallicity AGB stars experience a higher number of free neutrons, which results in higher values of both [hs/ls] and [$s$/Fe], while other stellar model features are secondary. To confirm this, we run one 2 M$_{\odot}$ Monash model of solar metallicity changing the metallicity during the post-processing only, i.e., keeping the same stellar structure such as the temperature and the amount of third dredge-up, and obtained the same results as those plotted in the figure. If we consider that the [$s$/Fe] ratio is affected by dilution, the prediction lines can be shifted to the left to cover all the data points. Therefore, we conclude that the slope of the line that fits the observations, as well as the spread around this line, is a product of the combination of the $s$-process intrinsic features dominated by the neutron exposure, in turn controlled mostly by the metallicity, and the effect of dilution due to binary mass transfer.

Recently, \citet{karinkuzhi2021} have proposed a new way to distinguish between CEMP-$s$, CEMP-$r$, and CEMP-$r/s$ stars based not only on the abundances of two elements (currently Ba and Eu), but on a set of abundances of heavy elements, by defining a signed rms distance, $d_{\rm{rms}}$, from the solar $r$-process pattern. According to these authors, the quantity $d_{\rm{rms}}$ is written as:

\begin{equation}
    d_{\rm{rms}}=\left\{\dfrac{1}{N}\sum_{x_i}[\log\epsilon(x_{i})_{\star}-\log\epsilon(x_{i})_{\rm{norm}(r,\star)}]^{2}\right\}^{1/2},
\end{equation}

\noindent where $N$ is the number of neutron-capture elements considered that make up the list $x_{i}$, $\log\epsilon(x_{i})_{\star}$ is the logarithmic abundance of the chemical species belonging to $x_{i}$, and

\begin{equation}
    \log\epsilon(x_{i})_{\textrm{norm}(r,\star)}=\log\epsilon(x_{i})_{r}+\left[\log\epsilon(\textrm{Eu})_{\star}-\log\epsilon(\textrm{Eu})_{r}\right],
\end{equation}

\noindent being $\log\epsilon(x_{i})_{r}$ the standard $r$-process abundance, given in Table B4 of \citeauthor{karinkuzhi2021}. Note that, by definition, $d_{\rm{rms}}=0$ for Eu; we refer the reader to Section 5 of that paper for a more detailed discussion. The quantity $d_{\rm{rms}}$ becomes higher for the more $s$-rich stars.

\begin{figure}
    \centering
    \includegraphics{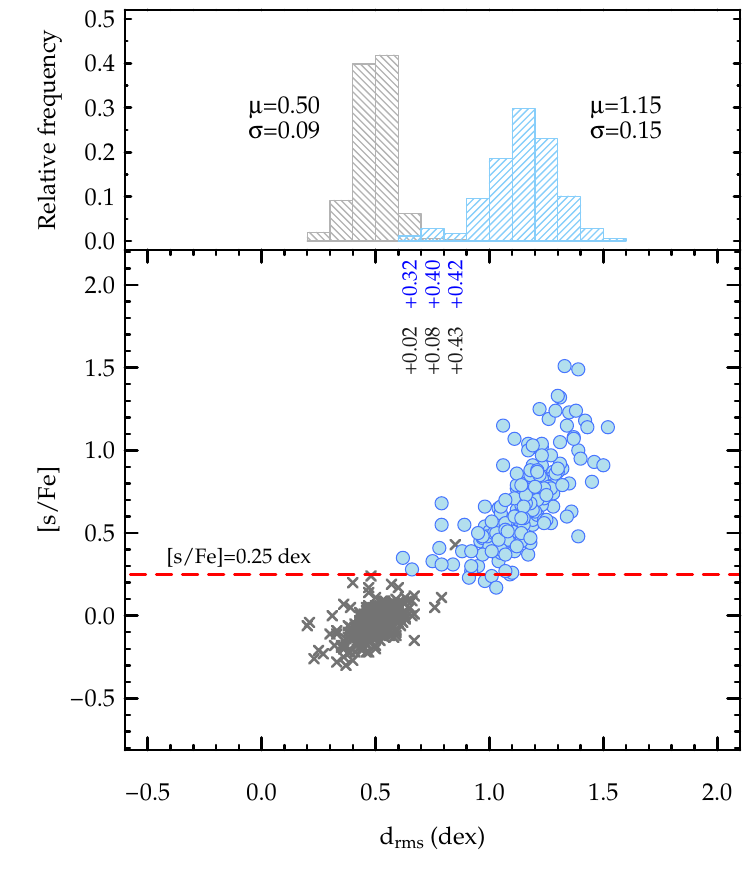}
    \caption{\textit{Upper panel}: distributions of the $d_{\rm{rms}}$ distance for field giant stars (grey bars) and Ba stars (blue bars), indicating the mean ($\mu$) and standard deviation ($\sigma$) for the two distributions. \textit{Lower panel}: field stars (grey crosses) and Ba stars (blue dots) in the diagram [$s$/Fe] against $d_{\rm{rms}}$; the numbers in vertical are the ratios [$s$/Fe] for the program Ba stars (blue) and field giant stars (grey) with $d_{\rm{rms}}$ between 0.6 and 0.9 dex. The red dashed line represents the limit assumed by \citet{decastro2016} to classify an object as a Ba star.}
    \label{fig:diagnostic}
\end{figure}

Since we have obtained Eu abundances for the program stars, we applied this method for the target objects in order to find an additional constraint that separates between Ba stars and field stars. As we pointed before, \citet{decastro2016} defined the condition [$s$/Fe] $\ge 0.25$ dex as a criterion to classify an object as a Ba star. We computed the $d_{\rm{rms}}$ distance for the Ba giant stars of this study, considering the elemental abundances of Rb, Sr, Y, Zr, Nb, Mo, Ru, La, Ce, Nd, and Sm; furthermore, we have also computed the $d_{\rm{rms}}$ values for the samples of field giant stars of \citet{luck2007} and \citet{mishenina2007}, taking into account the neutron-capture abundances provided by these authors. Whereas field giant stars show $d_{\rm{rms}}$ in the range of $0.2-0.9$ dex, with $\langle d_{\rm{rms}} \rangle_{\rm{field}} = 0.50$, for Ba stars we found $d_{\rm{rms}}$ between $0.6-1.6$ dex, with $\langle d_{\rm{rms}} \rangle_{\rm{Ba}} =1.15$ dex. In the upper panel of Fig. \ref{fig:diagnostic}, we show the distribution of $d_{\rm{rms}}$ for field giant stars and Ba stars. In the lower panel of the same figure, we plot the data set in the diagram [$s$/Fe] against $d_{\rm{rms}}$, where it is possible to notice that Ba stars (blue dots) behave differently of field stars (grey crosses); while Ba stars present a slope of $\sim 1.2$, field stars show a slope of $\sim 0.6$. The red dashed line in the figure represents the limit assumed by \citeauthor{decastro2016}. We note that the single grey cross with [$s$/Fe] $>0.25$ dex in Fig. \ref{fig:diagnostic} is the star HD 104979, which is actually a Ba star \citep{jorissen2019}. The Ba giant stars considered in the present study show both [$s$/Fe] $\geq 0.25$ dex and $d_{\rm{rms}}>0.6$ dex.

\section{Comparison to nucleosynthesis models}\label{sec:comparison}

\begin{figure*}
    \centering
    \includegraphics{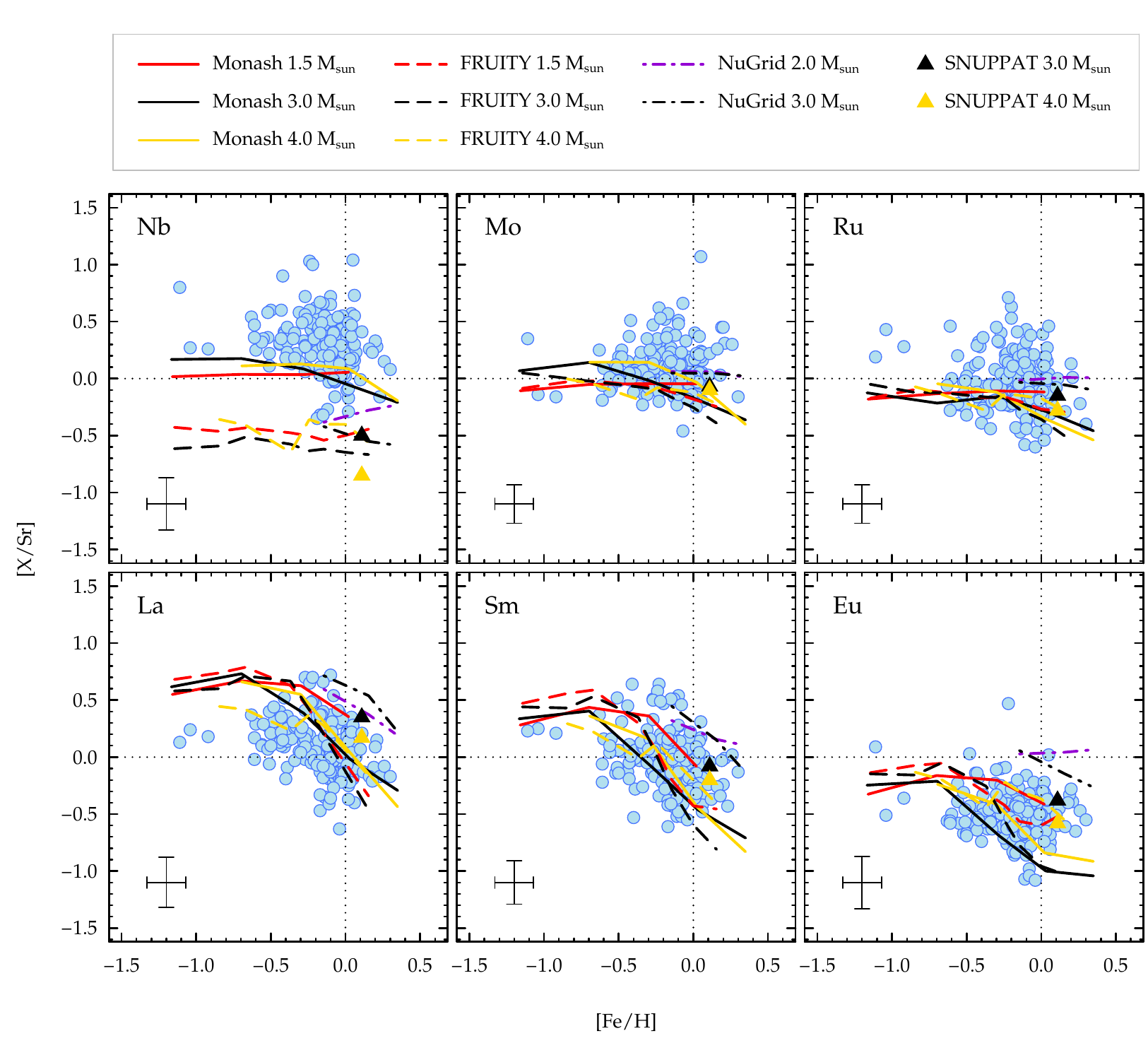}
    \caption{Comparison between observed (blue dots) and theoretical predictions for the [X/Sr] ratio from four different sets of the $s$-process models as calculated at the stellar surface at the end of the evolution. The Monash models are from \citet{fishlock2014, karakas2016, karakas2018}, with the standard choice of the partial mixing zone leading to the formation of the $^{13}$C pocket; the {\sc fruity} models are from the {\sc fruity} database \citep{cristallo2011}, not including rotation and with the standard choice of the $^{13}$C pocket; the NuGrid models are from \citet{battino2019}; the {\sc snuppat} models are from Yag\"ue L\'opez et al. (submitted, adopting here the models with overshoot parameter equal 0.14 leading to the formation of the $^{13}$C pocket). Note that only in the Monash models $^{93}$Zr is decayed into $^{93}$Nb and the final predicted $s$-process abundance for Nb is shown. Note that the observations and the models are respectively normalised to the solar photospheric abundances of \citet{grevesse1998} and the meteoritic abundances of \citet{asplund2009}. If the models were renormalised to the same as the data, changes would be of the order of $<0.1$ dex.}
    \label{fig:x-sr}
\end{figure*}

\begin{figure}
    \centering
    \includegraphics{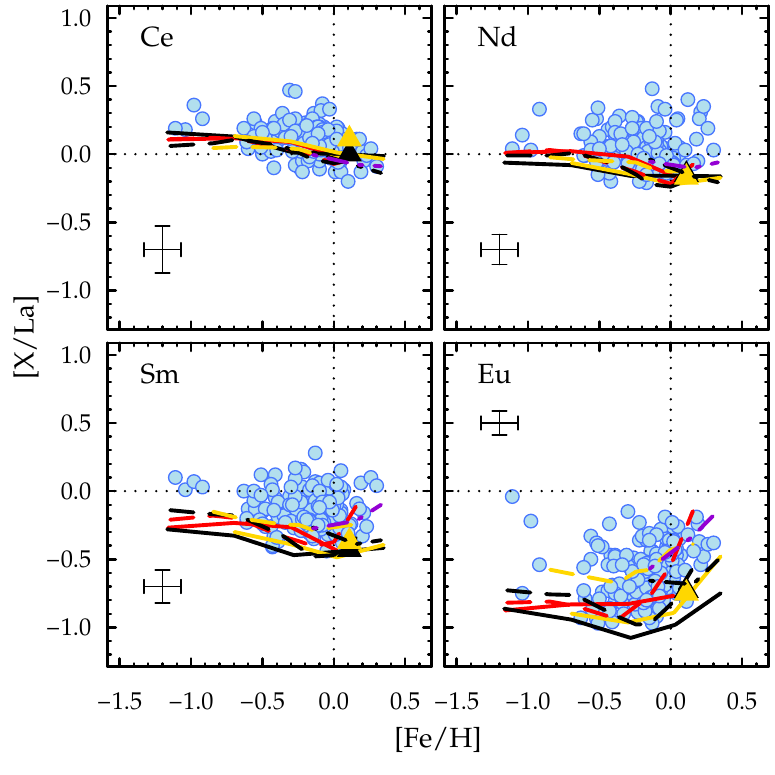}
    \caption{Comparison between observed and theoretical predictions for the [X/La]; the symbols have the same meaning as Fig. \ref{fig:x-sr}.}
    \label{fig:x-la}
\end{figure}

In Fig. \ref{fig:x-sr}, we present the abundance ratio of each element relative to Sr, taken as representative of the first $s$-process peak, and compare the data to AGB models from different databases, as listed in the figure caption. In Fig. \ref{fig:x-la}, we show the ratios of the elements belonging to the second $s$-process peak and Eu relative to La, another second $s$-process peak element. In other words, we have considered here the ratios [X/Sr] and [X/La], which allow us to determine the main features of the $s$-process relative abundance distribution and derive if such features are predicted by the models. We cannot directly compare the models to the [X/Fe] data because AGB model predictions do not include transfer of the AGB material in the binary system and its dilution in the convective envelope of the observed Ba giant stars, which would somewhat decrease the [X/Fe] ratios. Cseh et al. (in preparation) selected 28 Ba stars for which the mass has been independently determined \citep{jorissen2019} and compared the AGB models to each individual star. This analysis allowed them to determine quantitatively possible dilution factors for each AGB model, and verify if they are realistic. Here, instead, the only constraint considered in relation to [X/Fe] values is that we only plot compositions predicted at the end of the calculated AGB evolution that show [$s$/Fe] $\ge 0.25$ dex. While this extreme case implies no dilution on the secondary Ba star, which is unrealistic, we still use it for the general comparison presented here because predicted [$s$/Fe] ratios can change due to uncertainties related the amount of dredge-up and mass extent of the $^{13}$C pocket in the AGB models. In general, initial masses around 3 M$_{\odot}$ have the highest dredge-up efficiency and the largest $^{13}$C-pocket mass; therefore, these masses typically predict [X/Fe] ratios between 1 and 2 dex and allow for the highest dilution on the secondary Ba star.

The bottom panels of Fig. \ref{fig:x-sr}, where the second $s$-process peak elements La and, partly, Sm, and the $r$-process element Eu are divided by Sr, show the trend that was previously found by \citet{decastro2016} and demonstrated by \citet{cseh2018} to be consistent with the trend of AGB model predictions with metallicity (see also Fig. \ref{fig:s_mean}). Due to the primary nature of the $^{13}$C neutron source, the number of neutrons captured by iron seed decreases as iron increases, therefore the second $s$-process peak elements are more efficiently produced at lower metallicity and the ratios in these panels increase, as shown by the observational data. The spread present at any given metallicity may be due to different effects, from the stellar mass, to mixing processes, and investigation of individual stars of known mass will help us to disentangle the different possibilities (Cseh et al., in preparation). The [Eu/Sr] ratios are always negative, as expected since Sr and Eu are typical $s$- and $r$-process elements, respectively, in the Solar System (see Table \ref{tab:fraction}). Therefore, by definition, the $s$-process predicts stronger production of Sr, relative to Eu, relative to solar.

However, the behaviour of the ratio of the elements close to the first $s$-process peak (Nb, Mo, and Ru) relative to the first $s$-process peak element Sr shown in the top panels of Fig. \ref{fig:x-sr}, and the ratios of the elements close to the second $s$-process peak (Nd, Sm, and Ce, to a much lesser extent) relative to the second $s$-process peak element La shown in Fig. \ref{fig:x-la}, are puzzling. Even if the error bars are quite large, of the order of $\pm 0.2-0.3$ dex, it is not obvious why the data are on average higher than the models predictions. These trends are clearly at odds with all the $s$-process predictions, which reach at most +0.20 dex because the $s$-process predicted ratios behave as expected based on the contribution of the $s$- and $r$-processes to the different elements listed in Table \ref{tab:fraction}. Roughly speaking, by definition, the smaller the $s$-process contribution to a certain elements, the lower its $s$-process predicted ratio to Sr and La, which have some of the highest $s$-contribution. This behaviour reflects the neutron-capture cross sections of the isotopes involved, which are well known, rather than stellar model uncertainties, and it is not possible to modify it within the nuclear-physics framework of the $s$-process. Most individual stars can be reconciled with models when considering the associated uncertainties; however, there are some stars located too far from the model predictions to be matched within their error bars. Moreover, all the observed stellar abundance ratio distributions have medians above or at the upper range of the predicted values. For [Nb/Sr], the distribution is even more significantly distant from the models than the other ratios. Here below we discuss possible solutions to these problems.

\begin{figure*}
    \centering
    \includegraphics{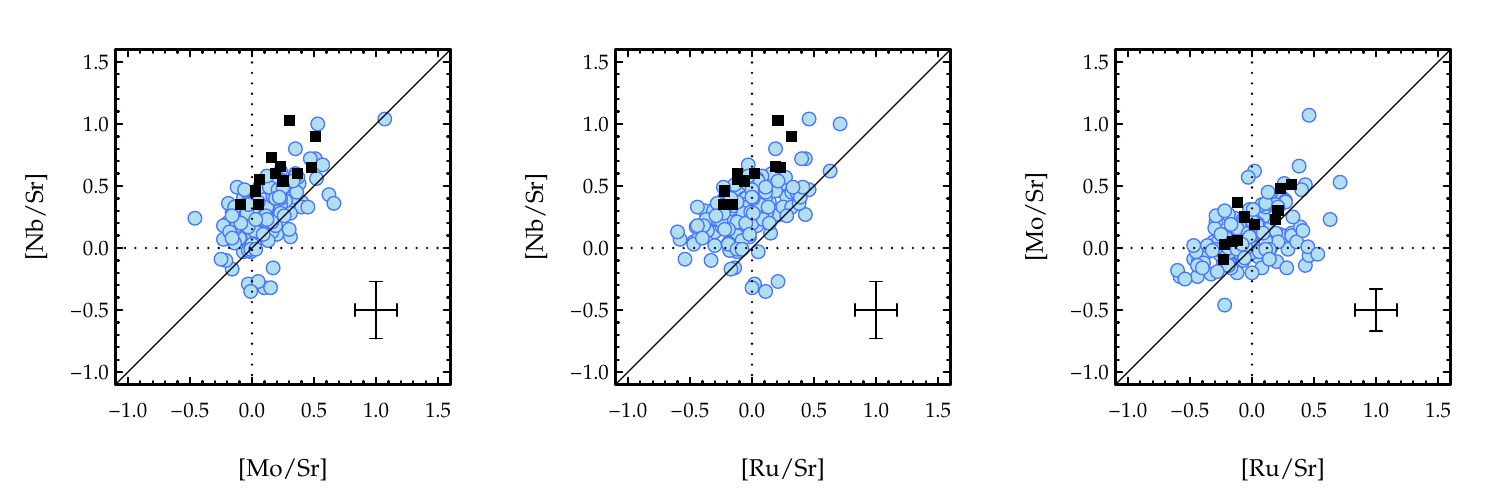}
    \caption{Correlation between the ratios [Nb,Mo,Ru/Sr] combining one with other. Black squares indicate stars that present [Nb/Fe] $\ge 1.5$ dex.}
    \label{fig:correlation}
\end{figure*}

One possibility is that there may be observational problems related to these elements that could produce the observed systematic shifts away from the model predictions. However, when we plot for example the [Nb,Mo,Ru/Sr] ratios against each other as in Fig. \ref{fig:correlation}, we see that the Nb, Mo, and Ru excesses relative to Sr are generally correlated, therefore it is not trivial to find a problem with the spectroscopic analysis that should apply in the same way to the three different elements.

The initial composition of the Ba star, which reflects the contribution of galactic chemical evolution to both the $s$- and $r$-processes, could be invoked to explain the mismatch. If we consider the abundances for field giant and dwarf stars (grey crosses in Figs. \ref{fig:s} and \ref{fig:sm-eu}), we find that there are a few stars with [Mo,Ru/Sr] and the [Ce,Nd,Sm,Eu/Sr] ratios significantly higher than solar, up to roughly 0.5 dex as observed also in some of the anomalous Ba stars. However, as explained at the start of this section, we need to select for the comparison AGB models that predict significant $s$-process [X/Fe] ratios, with at least [$s$/Fe] $\ge 0.25$ dex, otherwise the $s$-process signatures would not be observable in the Ba star envelopes after dilution. In other words, when considering Fig. \ref{fig:s}, we can see clearly that the absolute [X/Fe] abundances for field giant and dwarf stars are at least an order of magnitude lower than those of the Ba stars, showing that the transferred abundances to the observed Ba stars overwhelm their initial composition, and therefore they do not retain any memory of their initial abundance pattern. The effect of the initial composition may play a role for the [Eu/La] ratios, for which the observed value are mostly negative and for which, as shown in Fig. \ref{fig:sm-eu}, at least the [Eu/Fe] ratio can retain a memory of its initial value.

The other possibility is the activation of another nucleosynthetic process on top of the $s$-process in the companions of some of the observed stars. Intermediate neutron captures (the $i$-process) have been recently invoked to explain a variety of observational peculiarities \citep[from the companions of CEMP stars, to open cluster stars, and post-AGB stars, e.g.,][]{hampel2016, hampel2019, mishenina2015, choplin2020}. However, given the high neutron densities associated with the $i$-process, a signature should be production of Rb, at least at the same level as Sr. This is not observed in the Ba stars considered here \citep{roriz2021}.

\citet{mishenina2019b} analyzed a sample of disc stars and compared the observations of Mo and Ru with different sets of galactic chemical evolution (GCE) models. These authors reported that GCE models do not produce sufficient Mo and Ru to reproduce their observations of these elements in the Galaxy and therefore some nucleosynthetic contributions must be not yet included in GCE models, possibly a late source. \citet{kobayashi2020} investigated the galactic evolution of the elements from C to U. The abundances of Mo and Ru predicted in these models overproduce the observations when including the $\nu$-wind component from nascent neutron stars on top of the $s$-process component from AGB stars, and reproduce the observations when including the contribution of magneto-rotational supernovae on top of the $s$-process component from AGB stars and the $r$-process component from compact binary mergers (see their Fig. 32). Still, the issue remains open as $\nu$-wind and magneto-rotational supernovae yields are uncertain, and isotopic abundances need to be also investigated whenever possible.

Here, it is tempting to suggest that we may have identified observationally a possible additional galactic source of these elements in the AGB companions of some Ba stars, although there are no current stellar scenarios that could present a solution. One suggestion may be that in some of these AGB stars one or more late neutron fluxes may act on an already present, strongly enhanced $s$-process distribution, with features that would allow to shift a fraction of the abundances of the first (Sr, Y, and Zr) and second (Ba, La, and Ce) peak elements into elements with a few more protons (Nb, Mo, and Ru for the first peaks and Nd and Sm for the second peak), without overproducing Rb. The detailed features of the neutron flux that would allow this, and if they exist in nature, need to be investigated starting with parametric models of neutron captures.

\section{Conclusions}\label{sec:conclusions}

In the present study we have provided new elemental abundances for the neutron-capture species Sr, Nb, Mo, Ru, La, Sm, and Eu for a large sample of 180 Ba stars. The abundances, computed either by equivalent width measurements or spectral synthesis, were compared with field giant and dwarf stars in the same range of metallicity, taken from literature. Ba stars show a general enrichment in heavy elements, with respect to field stars; except for Eu, for which the Ba stars show a content similar to the field stars. We have also revised the available La abundances for the whole sample and we found [La/Fe] ratios up to $\sim$ 1.2 dex lower than abundances previously reported by \citet{decastro2016}, when we consider effects due to HFS. We examined the behavior of the data set in the planes [La/Fe] versus [Eu/Fe] and [Zr/Fe] versus [Nb/Fe]. In the [La/Fe] versus [Eu/Fe] plane, Ba stars lie in the $s$-rich region, close to CEMP-$s$ stars, whereas the present sample falls systematically below the region predicted by \citet{neyskens2015} in the [Zr/Fe] versus [Nb/Fe] plane. We examined the indexes [hs/ls] and [$s$/Fe] of the $s$-process and, as expected, the present sample shows a correlation between the indexes [hs/ls] and [$s$/Fe]. The target stars also show [$s$/Fe] $\geq 0.25$ dex, within error bars, with a rms distance (in dex) from the solar $r$-process pattern \citep{karinkuzhi2021} in the range $0.6<d_{\rm{rms}}<1.6$.

We compared the observed abundances to four different current sets of $s$-process nucleosynthesis models. Second-to-first $s$-process peak ratios and the ratios of the predominantly $r$-process element Eu to La can be interpreted within the models predictions. However, the [Nb,Mo,Ru/Sr] and [Ce,Nd,Sm/La] ratios show median values higher or at the upper limits of the model predictions, with stars higher than model predictions by more than the observational uncertainty. There is still no clear explanation for this behaviour, and neutron capture models need to be investigated to verify if this shift from the most abundant peak elements onto the following elements can be reproduced, and with which features of a neutron flux. Targeted observations and analysis of the stars most enhanced in Nb, Mo, and Ru for more elemental abundances will help shedding light on the features of the nucleosynthesis process needed to explain them.

\section*{Acknowledgements}
We are very grateful to the anonymous referee for his/her comments and suggestions that have improved the manuscript. M. P. R. acknowledges financial support from Coordena\c c\~ao de Aperfei\c coamento de Pessoal de N\'ivel Superior (CAPES). M.L. acknowledges the support of the Hungarian National Research, Development and Innovation Office (NKFI), grant KH\_18 130405. C. S. thanks the U.S. National Science Foundation for support under grant AST 1616040. C. S. also thanks the Brazilian Astronomical Society and Observat\'orio Nacional for travel support. A.I.K. acknowledges financial support from the Australian Research Council (discovery project 170100521) and from the Australian Research Council Centre of Excellence for All Sky Astrophysics in 3 Dimensions (ASTRO 3D), through project number CE170100013. N.A.D. acknowledges financial support by Russian Foundation for Basic Research (RFBR) according to the research projects 18-02-00554 and 18-52-06004. This work has made use of the {\sc vald} database, operated at Uppsala University, the Institute of Astronomy RAS in Moscow, and the University of Vienna.

\section*{Data Availability}
The data underlying this article will be shared on reasonable request to the corresponding author.






\appendix

\section{}

\begin{table}
\centering
\caption{Hyperfine structure components of the Nb\,{\sc i} lines at 4646 \AA, 5344 \AA, and 5350 \AA.}
\label{tab:app1}
	\begin{tabular}{cccccc} 
	\toprule
	$\lambda$ (\AA) &  $\log$ \textit{gf}  &  $\lambda$ (\AA) & $\log$ \textit{gf} &  $\lambda$ (\AA) & $\log$ \textit{gf}\\
	\midrule
    4606.691  &  -2.133  &  5344.035  &  -3.632  & 5350.637  &  -3.424 \\
    4606.712  &  -1.906  &  5344.056  &  -2.403  & 5350.657  &  -2.373 \\
    4606.713  &  -1.133  &  5344.079  &  -1.451  & 5350.674  &  -2.989 \\
    4606.729  &  -1.278  &  5344.083  &  -3.208  & 5350.679  &  -1.583 \\
    4606.732  &  -1.813  &  5344.102  &  -2.181  & 5350.691  &  -2.168 \\
    4606.746  &  -1.441  &  5344.122  &  -1.558  & 5350.707  &  -2.734 \\
    4606.751  &  -2.133  &  5344.126  &  -2.964  & 5350.711  &  -1.725 \\
    4606.751  &  -1.781  &  5344.142  &  -2.093  & 5350.721  &  -2.103 \\
    4606.762  &  -1.628  &  5344.160  &  -1.676  & 5350.734  &  -2.561 \\
    4606.763  &  -1.906  &  5344.163  &  -2.809  & 5350.737  &  -1.892 \\
    4606.769  &  -1.792  &  5344.176  &  -2.069  & 5350.746  &  -2.113 \\
    4606.775  &  -1.813  &  5344.191  &  -1.809  & 5350.757  &  -2.436 \\
    4606.776  &  -1.847  &  5344.195  &  -2.712  & 5350.759  &  -2.097 \\
    4606.784  &  -1.839  &  5344.205  &  -2.089  & 5350.766  &  -2.193 \\
    4606.787  &  -1.781  &  5344.218  &  -1.964  & 5350.775  &  -2.336 \\
    4606.790  &  -2.110  &  5344.220  &  -2.663  & 5350.776  &  -2.368 \\
    4606.797  &  -1.930  &  5344.228  &  -2.151  & 5350.780  &  -2.395 \\
    4606.797  &  -1.792  &  5344.238  &  -2.151  & 5350.789  &  -2.793 \\
    4606.801  &  -2.442  &  5344.240  &  -2.663  &           &         \\
    4606.807  &  -1.839  &  5344.245  &  -2.265  &           &         \\
    4606.808  &  -2.082  &  5344.253  &  -2.394  &           &         \\
    4606.810  &  -2.890  &  5344.254  &  -2.730  &           &         \\
    4606.813  &  -1.930  &  5344.257  &  -2.467  &           &         \\
    4606.814  &  -2.370  &  5344.262  &  -2.768  &           &         \\
    4606.816  &  -3.589  &            &          &           &         \\
    4606.818  &  -2.082  &            &          &           &         \\
    4606.820  &  -2.370  &            &          &           &         \\
	\bottomrule
	\end{tabular}
	
\end{table}

\begin{table}
\centering
\caption{Hyperfine structure components of the La\,{\sc ii} lines at 6320 \AA\ and 6774 \AA.}
\label{tab:app2}
	\begin{tabular}{cccc} 
	\toprule
	$\lambda$ (\AA) &  $\log$ \textit{gf} & $\lambda$ (\AA) &  $\log$ \textit{gf}\\
	\midrule
    6320.299 & -2.742 &  6774.132 & -3.762\\
    6320.311 & -2.919 &  6774.134 & -3.281\\
    6320.313 & -2.568 &  6774.143 & -3.281\\
    6320.330 & -4.669 &  6774.148 & -3.061\\
    6320.335 & -2.558 &  6774.162 & -3.061\\
    6320.342 & -2.742 &  6774.166 & -3.912\\
    6320.357 & -3.249 &  6774.171 & -2.960\\
    6320.365 & -2.717 &  6774.189 & -2.960\\
    6320.374 & -2.568 &  6774.194 & -3.278\\
    6320.392 & -2.543 &  6774.201 & -2.935\\
    6320.414 & -2.558 &  6774.224 & -2.935\\
    6320.435 & -2.147 &  6774.231 & -2.889\\
    6320.462 & -2.717 &  6774.240 & -2.987\\
             &        &  6774.267 & -2.987\\
             &        &  6774.276 & -2.603\\
             &        &  6774.286 & -3.182\\
             &        &  6774.317 & -3.182\\
             &        &  6774.328 & -2.375\\
	\bottomrule
	\end{tabular}
\end{table}

\clearpage
\onecolumn

\begin{longtable}{lcccccccc}
\caption{New elemental abundances obtained in this study for the program Ba stars.} \label{tab:app3} \\

\toprule
\multicolumn{1}{l}{Star} & 
\multicolumn{1}{c}{[Fe/H]} & 
\multicolumn{1}{c}{[Sr/Fe]} & 
\multicolumn{1}{c}{[Nb/Fe]} & 
\multicolumn{1}{c}{[Mo/Fe]} & 
\multicolumn{1}{c}{[Ru/Fe]} &
\multicolumn{1}{c}{[La/Fe]} & 
\multicolumn{1}{c}{[Sm/Fe]} & 
\multicolumn{1}{c}{[Eu/Fe]} \\
\midrule
\endfirsthead

\multicolumn{9}{c}%
{{\tablename\ \thetable{} -- Continued from previous page}} \\
\midrule 
\multicolumn{1}{l}{Star} & 
\multicolumn{1}{c}{[Fe/H]} & 
\multicolumn{1}{c}{[Sr/Fe]} & 
\multicolumn{1}{c}{[Nb/Fe]} & 
\multicolumn{1}{c}{[Mo/Fe]} & 
\multicolumn{1}{c}{[Ru/Fe]} &
\multicolumn{1}{c}{[La/Fe]} & 
\multicolumn{1}{c}{[Sm/Fe]} & 
\multicolumn{1}{c}{[Eu/Fe]} \\
\midrule
\endhead

\midrule
\multicolumn{9}{r}{{\textit{Continued on next page}}} \\
\endfoot

\hline
\endlastfoot

BD$-08^{\circ}$ 3194      & $-$0.10 $\pm$ 0.16 & 0.66 $\pm$ 0.28 & 1.38 $\pm$ 0.25 & 1.17 $\pm$ 0.24 & 1.09 $\pm$ 0.25 & 1.38 $\pm$ 0.21 & 1.20 $\pm$ 0.23 & 0.50 $\pm$ 0.17\\
BD$-09^{\circ}$ 4337      & $-$0.24 $\pm$ 0.21 & 1.35 $\pm$ 0.28 & 2.38 $\pm$ 0.25 & 1.65 $\pm$ 0.25 & 1.56 $\pm$ 0.25 & 1.54 $\pm$ 0.23 & 1.26 $\pm$ 0.25 & 0.81 $\pm$ 0.17\\
BD$-14^{\circ}$ 2678      & $+$0.01 $\pm$ 0.12 & 0.69  & 1.06  & 1.09 $\pm$ 0.24 & 0.83  & 0.91 $\pm$ 0.20 & 0.72 $\pm$ 0.23 & 0.25 $\pm$ 0.18\\
CD$-27^{\circ}$ 2233      & $-$0.25 $\pm$ 0.18 & 0.83 $\pm$ 0.27 & 1.10 $\pm$ 0.25 & 1.10 $\pm$ 0.25 & 0.81 $\pm$ 0.26 & 0.93 $\pm$ 0.20 & 0.82 $\pm$ 0.24 & 0.23 $\pm$ 0.17\\
CD$-29^{\circ}$ 8822      & $+$0.04 $\pm$ 0.15 & 0.91 $\pm$ 0.26 & 1.03  & 0.91 $\pm$ 0.24 & 1.06  & 0.96 $\pm$ 0.21 & 0.77 $\pm$ 0.23 & 0.21 $\pm$ 0.18\\
CD$-30^{\circ}$ 8774      & $-$0.11 $\pm$ 0.14 & 0.51  & 0.48  & 0.50 $\pm$ 0.25 & 0.39  & 0.32 $\pm$ 0.21 & 0.35 $\pm$ 0.24 &$-$0.18 $\pm$ 0.17\\
CD$-38^{\circ}$ 585       & $-$0.52 $\pm$ 0.09 & 1.02 $\pm$ 0.28 & 1.38 $\pm$ 0.26 & 1.06 $\pm$ 0.25 & 0.89  & 1.29 $\pm$ 0.21 & 1.08 $\pm$ 0.24 & 0.35 $\pm$ 0.17\\
CD$-42^{\circ}$ 2048      & $-$0.23 $\pm$ 0.16 & 0.99 $\pm$ 0.32 & 1.26 $\pm$ 0.26 & 1.10 $\pm$ 0.26 & 1.21 $\pm$ 0.30 & 0.98 $\pm$ 0.21 & 1.14 $\pm$ 0.25 & 0.23 $\pm$ 0.19\\
CD$-53^{\circ}$ 8144      & $-$0.19 $\pm$ 0.15 & 0.58  & 1.14  & 1.10 $\pm$ 0.25 & 0.84 $\pm$ 0.26 & 0.94 $\pm$ 0.20 & 0.66 $\pm$ 0.24 & 0.17 $\pm$ 0.17\\
CD$-61^{\circ}$ 1941      & $-$0.20 $\pm$ 0.14 & 0.46 $\pm$ 0.28 & 1.08  & 0.69 $\pm$ 0.25 & 1.09  & 0.92 $\pm$ 0.20 & 0.93 $\pm$ 0.24 & 0.19 $\pm$ 0.17\\

CPD$-62^{\circ}$ 1013     & $-$0.08 $\pm$ 0.14 & 0.84 $\pm$ 0.26 & $...$ & 0.93 $\pm$ 0.24 & 1.01  & 0.74 $\pm$ 0.20 & 0.50 $\pm$ 0.23 & 0.06 $\pm$ 0.18\\
CPD$-64^{\circ}$ 4333     & $-$0.10 $\pm$ 0.18 & 0.95 $\pm$ 0.28 & 1.60 $\pm$ 0.26 & 1.43 $\pm$ 0.26 & 1.18 $\pm$ 0.26 & 1.32 $\pm$ 0.20 & 1.22 $\pm$ 0.24 & 0.40 $\pm$ 0.17\\
HD 4084         & $-$0.42 $\pm$ 0.15 & 0.65 $\pm$ 0.30 & 1.55 $\pm$ 0.26 & 1.16 $\pm$ 0.26 & 0.97 $\pm$ 0.25 & 0.94 $\pm$ 0.20 & 0.61 $\pm$ 0.24 & 0.47 $\pm$ 0.17\\
HD 5424         & $-$0.41 $\pm$ 0.18 & 0.92 $\pm$ 0.28 & 1.33  & 1.02 $\pm$ 0.24 & 1.20  & 1.44 $\pm$ 0.21 & 1.56 $\pm$ 0.24 & 0.48 $\pm$ 0.17\\
HD 5825         & $-$0.48 $\pm$ 0.08 & 0.66  & $...$ & 0.91 $\pm$ 0.25 & 0.75  & 0.97 $\pm$ 0.20 & 0.71 $\pm$ 0.23 & 0.23 $\pm$ 0.18\\
HD 15589        & $-$0.27 $\pm$ 0.15 & 0.70 $\pm$ 0.28 & 1.42 $\pm$ 0.26 & 1.17 $\pm$ 0.25 & 1.10 $\pm$ 0.25 & 1.40 $\pm$ 0.20 & 1.21 $\pm$ 0.24 & 0.60 $\pm$ 0.17\\
HD 20394        & $-$0.22 $\pm$ 0.12 & 0.84 $\pm$ 0.27 & 1.25 $\pm$ 0.27 & 1.17 $\pm$ 0.24 & 0.95 $\pm$ 0.22 & 1.33 $\pm$ 0.20 & 1.04 $\pm$ 0.23 & 0.50 $\pm$ 0.18\\
HD 21989        & $-$0.14 $\pm$ 0.17 & 0.36 $\pm$ 0.32 & 0.47 $\pm$ 0.27 & 0.56 $\pm$ 0.26 & 0.26 $\pm$ 0.30 & 0.51 $\pm$ 0.21 & 0.50 $\pm$ 0.24 & 0.01 $\pm$ 0.19\\
HD 22285        & $-$0.60 $\pm$ 0.13 & 0.95 $\pm$ 0.27 & 1.20  & 1.04 $\pm$ 0.25 & 0.95  & 1.24 $\pm$ 0.20 & 1.11 $\pm$ 0.24 & 0.45 $\pm$ 0.17\\
HD 22772        & $-$0.17 $\pm$ 0.13 & 0.46 $\pm$ 0.30 & 0.79  & 0.86 $\pm$ 0.26 & 0.71 $\pm$ 0.27 & 0.85 $\pm$ 0.20 & 0.54 $\pm$ 0.23 & 0.15 $\pm$ 0.17\\

HD 24035        & $-$0.23 $\pm$ 0.15 & 0.97 $\pm$ 0.28 & 1.40 $\pm$ 0.25 & 1.59 $\pm$ 0.24 & 0.99  & 1.45 $\pm$ 0.21 & 1.61 $\pm$ 0.24 & 0.56 $\pm$ 0.17\\
HD 29370        & $-$0.25 $\pm$ 0.16 & 0.76 $\pm$ 0.31 & 0.85  & 1.07 $\pm$ 0.24 & 0.74 $\pm$ 0.25 & 0.88 $\pm$ 0.20 & 0.75 $\pm$ 0.23 & 0.05 $\pm$ 0.17\\
HD 29685        & $-$0.07 $\pm$ 0.14 & 0.44  & 0.28  & 0.61 $\pm$ 0.24 & 0.30  & 0.51 $\pm$ 0.20 & 0.40 $\pm$ 0.24 &$-$0.14 $\pm$ 0.17\\
HD 30240        & $+$0.02 $\pm$ 0.15 & 0.64 $\pm$ 0.26 & 0.35  & 0.61 $\pm$ 0.23 & 0.66  & 0.60 $\pm$ 0.20 & 0.44 $\pm$ 0.23 &$-$0.04 $\pm$ 0.18\\
HD 30554        & $-$0.12 $\pm$ 0.14 & 0.40 $\pm$ 0.28 & 0.92  & 0.78 $\pm$ 0.25 & 0.67 $\pm$ 0.26 & 0.68 $\pm$ 0.20 & 0.56 $\pm$ 0.23 &$-$0.01 $\pm$ 0.17\\
HD 32712        & $-$0.24 $\pm$ 0.16 & 0.30  & 0.90 $\pm$ 0.26 & 0.65 $\pm$ 0.25 & 0.46 $\pm$ 0.30 & 0.95 $\pm$ 0.21 & 0.88 $\pm$ 0.23 & 0.21 $\pm$ 0.19\\
HD 32901        & $-$0.44 $\pm$ 0.14 & 0.27 $\pm$ 0.31 & 0.57 $\pm$ 0.26 & 0.42 $\pm$ 0.25 & 0.39 $\pm$ 0.30 & 0.67 $\pm$ 0.20 & 0.72 $\pm$ 0.24 & 0.13 $\pm$ 0.19\\
HD 35993        & $-$0.05 $\pm$ 0.12 & 0.80 $\pm$ 0.26 & 1.22 $\pm$ 0.26 & 0.95 $\pm$ 0.23 & 0.84 $\pm$ 0.23 & 1.10 $\pm$ 0.20 & 0.88 $\pm$ 0.23 & 0.30 $\pm$ 0.18\\
HD 36650        & $-$0.28 $\pm$ 0.13 & 0.51  & 0.84 $\pm$ 0.25 & 0.63 $\pm$ 0.24 & 0.83 $\pm$ 0.26 & 0.64 $\pm$ 0.20 & 0.37 $\pm$ 0.24 & 0.08 $\pm$ 0.17\\
HD 38488        & $+$0.05 $\pm$ 0.10 & 0.61 $\pm$ 0.31 & 0.89 $\pm$ 0.25 & 0.85 $\pm$ 0.26 & 0.47 $\pm$ 0.30 & 0.63 $\pm$ 0.21 & 0.78 $\pm$ 0.25 & 0.08 $\pm$ 0.19\\

HD 40430        & $-$0.23 $\pm$ 0.13 & 0.62  & 0.88  & 0.59 $\pm$ 0.24 & 0.69 $\pm$ 0.25 & 0.70 $\pm$ 0.20 & 0.54 $\pm$ 0.24 & 0.14 $\pm$ 0.17\\
HD 43389        & $-$0.50 $\pm$ 0.17 & 1.15 $\pm$ 0.32 & 1.75 $\pm$ 0.25 & 1.34 $\pm$ 0.25 & 1.17 $\pm$ 0.30 & 1.36 $\pm$ 0.22 & 1.21 $\pm$ 0.24 & 0.48 $\pm$ 0.19\\
HD 51959        & $-$0.10 $\pm$ 0.15 & 0.68 $\pm$ 0.26 & 1.11 $\pm$ 0.26 & 0.86 $\pm$ 0.23 & 0.66 $\pm$ 0.22 & 0.82 $\pm$ 0.20 & 0.68 $\pm$ 0.23 & 0.17 $\pm$ 0.18\\
HD 58368        & $+$0.04 $\pm$ 0.14 & 0.45 $\pm$ 0.27 & 1.02 $\pm$ 0.25 & 0.82 $\pm$ 0.23 & 0.72 $\pm$ 0.23 & 0.80 $\pm$ 0.21 & 0.59 $\pm$ 0.25 & 0.09 $\pm$ 0.18\\
HD 59852        & $-$0.22 $\pm$ 0.10 & 0.32  & $...$ & 0.34 $\pm$ 0.24 & 0.46  & 0.27 $\pm$ 0.21 & 0.30 $\pm$ 0.24 &$-$0.09 $\pm$ 0.18\\
HD 61332        & $+$0.07 $\pm$ 0.13 & 0.39 $\pm$ 0.27 & 0.62 $\pm$ 0.25 & 0.33 $\pm$ 0.25 & 0.16 $\pm$ 0.25 & 0.39 $\pm$ 0.22 & 0.40 $\pm$ 0.26 &$-$0.08 $\pm$ 0.17\\
HD 64425        & $+$0.06 $\pm$ 0.16 & 0.70 $\pm$ 0.30 & 1.25 $\pm$ 0.25 & 0.93 $\pm$ 0.24 & 0.66 $\pm$ 0.25 & 0.93 $\pm$ 0.21 & 0.56 $\pm$ 0.23 & 0.18 $\pm$ 0.17\\
HD 66291        & $-$0.31 $\pm$ 0.15 & 0.34 $\pm$ 0.32 & 0.92 $\pm$ 0.26 & 0.69 $\pm$ 0.25 & 0.42 $\pm$ 0.31 & 0.62 $\pm$ 0.21 & 0.58 $\pm$ 0.24 & 0.11 $\pm$ 0.19\\
HD 67036        & $-$0.41 $\pm$ 0.13 & 0.74 $\pm$ 0.31 & 1.13 $\pm$ 0.25 & 0.81 $\pm$ 0.25 & 0.51 $\pm$ 0.30 & 0.83 $\pm$ 0.21 & 0.78 $\pm$ 0.25 & 0.14 $\pm$ 0.19\\
HD 71458        & $-$0.03 $\pm$ 0.10 & 0.48 $\pm$ 0.31 & 0.91 $\pm$ 0.25 & 0.59 $\pm$ 0.25 & 0.44 $\pm$ 0.30 & 0.63 $\pm$ 0.21 & 0.65 $\pm$ 0.25 & 0.12 $\pm$ 0.19\\

HD 74950        & $-$0.21 $\pm$ 0.13 & 0.50 $\pm$ 0.32 & 0.73 $\pm$ 0.25 & 0.52 $\pm$ 0.25 & 0.40 $\pm$ 0.30 & 0.52 $\pm$ 0.21 & 0.66 $\pm$ 0.25 & 0.02 $\pm$ 0.19\\
HD 82221        & $-$0.21 $\pm$ 0.18 & 0.72 $\pm$ 0.32 & 1.03 $\pm$ 0.25 & 0.80 $\pm$ 0.25 & 0.48 $\pm$ 0.30 & 0.74 $\pm$ 0.21 & 0.58 $\pm$ 0.24 & 0.09 $\pm$ 0.19\\
HD 83548        & $+$0.03 $\pm$ 0.14 & 0.35 $\pm$ 0.27 & 0.81  & 0.47 $\pm$ 0.23 & 0.54 $\pm$ 0.25 & 0.58 $\pm$ 0.21 & 0.35 $\pm$ 0.23 & 0.01 $\pm$ 0.18\\
HD 84610        & $+$0.00 $\pm$ 0.14 & 0.55 $\pm$ 0.28 & 0.92 $\pm$ 0.25 & 0.59 $\pm$ 0.24 & 0.57 $\pm$ 0.25 & 0.51 $\pm$ 0.21 & 0.47 $\pm$ 0.24 & 0.07 $\pm$ 0.17\\
HD 84678        & $-$0.13 $\pm$ 0.16 & 1.03 $\pm$ 0.32 & 1.36 $\pm$ 0.25 & 1.24 $\pm$ 0.25 & 0.74 $\pm$ 0.31 & 1.26 $\pm$ 0.22 & 1.54 $\pm$ 0.24 & 0.29 $\pm$ 0.19\\
HD 88035        & $-$0.10 $\pm$ 0.18 & 0.78 $\pm$ 0.28 & 1.23 $\pm$ 0.25 & 0.88 $\pm$ 0.24 & 1.10 $\pm$ 0.26 & 1.04 $\pm$ 0.20 & 0.87 $\pm$ 0.24 & 0.19 $\pm$ 0.17\\
HD 88562        & $-$0.27 $\pm$ 0.15 & 0.46 $\pm$ 0.32 & 1.02 $\pm$ 0.26 & 0.66 $\pm$ 0.25 & 0.50 $\pm$ 0.30 & 0.87 $\pm$ 0.21 & 0.96 $\pm$ 0.25 & 0.21 $\pm$ 0.19\\
HD 89175        & $-$0.55 $\pm$ 0.13 & 1.18 $\pm$ 0.29 & 1.53  & 1.09 $\pm$ 0.24 & 0.95 $\pm$ 0.26 & 1.47 $\pm$ 0.20 & 1.34 $\pm$ 0.24 & 0.60 $\pm$ 0.17\\
HD 91208        & $+$0.05 $\pm$ 0.14 & 0.88 $\pm$ 0.27 & 0.98  & 0.78 $\pm$ 0.23 & 0.72 $\pm$ 0.23 & 0.69 $\pm$ 0.21 & 0.55 $\pm$ 0.23 & 0.12 $\pm$ 0.18\\
HD 91979        & $-$0.11 $\pm$ 0.12 & 0.85 $\pm$ 0.30 & 1.03 $\pm$ 0.25 & 0.90 $\pm$ 0.24 & 0.61 $\pm$ 0.26 & 0.80 $\pm$ 0.21 & 0.58 $\pm$ 0.24 & 0.07 $\pm$ 0.17\\

HD 92626        & $-$0.15 $\pm$ 0.22 & 0.98 $\pm$ 0.28 & 1.64 $\pm$ 0.25 & 1.21 $\pm$ 0.24 & 1.17  & 1.64 $\pm$ 0.22 & 1.49 $\pm$ 0.25 & 0.71 $\pm$ 0.17\\
HD 105902       & $-$0.18 $\pm$ 0.17 & 1.20 $\pm$ 0.30 & 1.66 $\pm$ 0.27 & 1.23 $\pm$ 0.24 & 0.98 $\pm$ 0.25 & 1.22 $\pm$ 0.21 & 1.13 $\pm$ 0.24 & 0.34 $\pm$ 0.17\\
HD 107264       & $-$0.19 $\pm$ 0.17 & 0.93  & 1.38  & 0.98 $\pm$ 0.26 & 0.82 $\pm$ 0.30 & 0.86 $\pm$ 0.21 & 0.81 $\pm$ 0.25 & 0.19 $\pm$ 0.19\\
HD 107541       & $-$0.63 $\pm$ 0.11 & 1.31 $\pm$ 0.26 & 1.85 $\pm$ 0.27 & 1.56 $\pm$ 0.24 & 1.25  & 1.68 $\pm$ 0.21 & 1.68 $\pm$ 0.24 & 0.75 $\pm$ 0.18\\
HD 110483       & $-$0.04 $\pm$ 0.14 & 0.62 $\pm$ 0.30 & 1.16 $\pm$ 0.25 & 0.81 $\pm$ 0.24 & 0.67 $\pm$ 0.25 & 1.01 $\pm$ 0.21 & 0.69 $\pm$ 0.24 & 0.28 $\pm$ 0.17\\
HD 110591       & $-$0.56 $\pm$ 0.12 & 0.52  & $...$ & 0.41 $\pm$ 0.24 & 0.72  & 0.63 $\pm$ 0.20 & 0.56 $\pm$ 0.23 & 0.13 $\pm$ 0.17\\
HD 111315       & $+$0.04 $\pm$ 0.09 & 0.54 $\pm$ 0.29 & 0.68  & 0.50 $\pm$ 0.24 & 0.48 $\pm$ 0.26 & 0.46 $\pm$ 0.20 & 0.30 $\pm$ 0.24 &$-$0.01 $\pm$ 0.17\\
HD 113291       & $-$0.02 $\pm$ 0.16 & 0.71 $\pm$ 0.28 & 1.26 $\pm$ 0.25 & 0.87 $\pm$ 0.24 & 0.67 $\pm$ 0.25 & 1.04 $\pm$ 0.20 & 0.93 $\pm$ 0.24 & 0.33 $\pm$ 0.17\\
HD 116869       & $-$0.36 $\pm$ 0.12 & 0.62 $\pm$ 0.28 & 1.02  & 0.55 $\pm$ 0.24 & 0.56 $\pm$ 0.26 & 0.85 $\pm$ 0.20 & 0.73 $\pm$ 0.24 & 0.17 $\pm$ 0.17\\
HD 119185       & $-$0.43 $\pm$ 0.10 & 0.31  & $...$ & 0.30 $\pm$ 0.24 & 0.60 $\pm$ 0.25 & 0.41 $\pm$ 0.20 & 0.42 $\pm$ 0.24 & 0.17 $\pm$ 0.17\\

HD 120571       & $-$0.39 $\pm$ 0.09 & 0.45  & 0.59 $\pm$ 0.25 & 0.35 $\pm$ 0.25 & 0.38 $\pm$ 0.31 & 0.59 $\pm$ 0.20 & 0.51 $\pm$ 0.23 & 0.12 $\pm$ 0.19\\
HD 120620       & $-$0.14 $\pm$ 0.18 & 1.13 $\pm$ 0.26 & 1.68 $\pm$ 0.26 & 1.19 $\pm$ 0.23 & 1.01 $\pm$ 0.23 & 1.41 $\pm$ 0.20 & 1.06 $\pm$ 0.23 & 0.61 $\pm$ 0.18\\
HD 122687       & $-$0.07 $\pm$ 0.13 & 0.87 $\pm$ 0.26 & 1.11 $\pm$ 0.26 & 0.41 $\pm$ 0.23 & 0.65 $\pm$ 0.23 & 0.91 $\pm$ 0.20 & 0.70 $\pm$ 0.22 & 0.18 $\pm$ 0.18\\
HD 123396       & $-$1.04 $\pm$ 0.13 & 0.87  & 1.14  & 0.73 $\pm$ 0.25 & 1.30  & 1.11 $\pm$ 0.20 & 1.12 $\pm$ 0.23 & 0.36 $\pm$ 0.19\\
HD 123701       & $-$0.44 $\pm$ 0.09 & 1.05 $\pm$ 0.26 & 1.36  & 1.07 $\pm$ 0.24 & 1.02 $\pm$ 0.23 & 1.18 $\pm$ 0.20 & 0.92 $\pm$ 0.23 & 0.45 $\pm$ 0.18\\
HD 123949       & $-$0.09 $\pm$ 0.18 & 0.85 $\pm$ 0.31 & 1.36 $\pm$ 0.25 & 1.01 $\pm$ 0.26 & 0.71 $\pm$ 0.31 & 1.16 $\pm$ 0.21 & 0.99 $\pm$ 0.24 & 0.38 $\pm$ 0.19\\
HD 126313       & $-$0.10 $\pm$ 0.16 & 0.86 $\pm$ 0.30 & 1.12 $\pm$ 0.25 & 0.85 $\pm$ 0.24 & 0.90 $\pm$ 0.26 & 0.91 $\pm$ 0.20 & 0.72 $\pm$ 0.24 & 0.21 $\pm$ 0.17\\
HD 130255       & $-$1.11 $\pm$ 0.11 & 0.19  & 0.99  & 0.54 $\pm$ 0.25 & 0.38 $\pm$ 0.30 & 0.32 $\pm$ 0.20 & 0.42 $\pm$ 0.24 & 0.28 $\pm$ 0.19\\
HD 131670       & $-$0.04 $\pm$ 0.15 & 0.61 $\pm$ 0.28 & 0.95 $\pm$ 0.25 & 0.62 $\pm$ 0.24 & 0.62 $\pm$ 0.26 & 0.58 $\pm$ 0.21 & 0.54 $\pm$ 0.24 & 0.00 $\pm$ 0.17\\
HD 136636       & $-$0.04 $\pm$ 0.18 & 0.81 $\pm$ 0.27 & 1.14 $\pm$ 0.25 & 0.81 $\pm$ 0.24 & 0.66 $\pm$ 0.25 & 0.92 $\pm$ 0.20 & 0.77 $\pm$ 0.24 & 0.23 $\pm$ 0.17\\

HD 142751       & $-$0.10 $\pm$ 0.13 & 0.70 $\pm$ 0.32 & 1.00 $\pm$ 0.25 & 0.70 $\pm$ 0.25 & 0.32 $\pm$ 0.30 & 0.72 $\pm$ 0.22 & 0.65 $\pm$ 0.24 & 0.04 $\pm$ 0.19\\
HD 143899       & $-$0.27 $\pm$ 0.12 & 0.66  & 0.95  & 0.60 $\pm$ 0.23 & 0.50 $\pm$ 0.24 & 0.64 $\pm$ 0.20 & 0.54 $\pm$ 0.23 & 0.05 $\pm$ 0.18\\
HD 147884       & $-$0.09 $\pm$ 0.15 & 0.85 $\pm$ 0.27 & 1.07  & 0.81 $\pm$ 0.23 & 0.73 $\pm$ 0.24 & 0.80 $\pm$ 0.20 & 0.52 $\pm$ 0.23 & 0.13 $\pm$ 0.18\\
HD 148177       & $-$0.15 $\pm$ 0.15 & 0.53 $\pm$ 0.35 & 0.91 $\pm$ 0.25 & 0.80 $\pm$ 0.25 & 0.34 $\pm$ 0.30 & 0.45 $\pm$ 0.21 & 0.39 $\pm$ 0.25 &$-$0.10 $\pm$ 0.19\\
HD 154430       & $-$0.36 $\pm$ 0.19 & 0.87 $\pm$ 0.32 & 1.00 $\pm$ 0.26 & 1.03 $\pm$ 0.25 & 0.57 $\pm$ 0.31 & 1.03 $\pm$ 0.22 & 0.84 $\pm$ 0.24 & 0.24 $\pm$ 0.19\\
HD 162806       & $-$0.26 $\pm$ 0.17 & 0.71 $\pm$ 0.35 & 0.98 $\pm$ 0.25 & 0.80 $\pm$ 0.25 & 0.48 $\pm$ 0.30 & 0.77 $\pm$ 0.21 & 0.54 $\pm$ 0.24 & 0.10 $\pm$ 0.19\\
HD 168214       & $-$0.08 $\pm$ 0.10 & 1.06 $\pm$ 0.27 & 1.47 $\pm$ 0.26 & 1.12 $\pm$ 0.23 & 0.88 $\pm$ 0.24 & 0.84 $\pm$ 0.20 & 0.63 $\pm$ 0.23 & 0.08 $\pm$ 0.18\\
HD 168560       & $-$0.13 $\pm$ 0.13 & 0.43  & 0.60 $\pm$ 0.25 & 0.51 $\pm$ 0.26 & 0.19 $\pm$ 0.30 & 0.40 $\pm$ 0.21 & 0.46 $\pm$ 0.25 &$-$0.04 $\pm$ 0.19\\
HD 168791       & $-$0.23 $\pm$ 0.17 & 0.86  & 1.35 $\pm$ 0.25 & 0.74 $\pm$ 0.25 & 0.63 $\pm$ 0.30 & 0.81 $\pm$ 0.21 & 0.52 $\pm$ 0.24 & 0.27 $\pm$ 0.19\\
HD 176105       & $-$0.14 $\pm$ 0.12 & 0.72 $\pm$ 0.34 & 0.77 $\pm$ 0.25 & 0.63 $\pm$ 0.26 & 0.25 $\pm$ 0.31 & 0.37 $\pm$ 0.21 & 0.29 $\pm$ 0.24 &$-$0.07 $\pm$ 0.19\\

HD 177192       & $-$0.17 $\pm$ 0.20 & 0.81  & 0.89 $\pm$ 0.26 & 0.71 $\pm$ 0.24 & 0.38  & 0.34 $\pm$ 0.21 & 0.20 $\pm$ 0.24 & $...$\\
HD 180996       & $+$0.06 $\pm$ 0.15 & 0.76  & 0.92 $\pm$ 0.25 & 0.56 $\pm$ 0.24 & 0.64  & 0.36 $\pm$ 0.20 & 0.34 $\pm$ 0.25 & 0.01 $\pm$ 0.17\\
HD 182300       & $+$0.06 $\pm$ 0.16 & 0.81 $\pm$ 0.26 & 1.11 $\pm$ 0.25 & 0.86 $\pm$ 0.23 & 0.50 $\pm$ 0.24 & 0.87 $\pm$ 0.20 & 0.63 $\pm$ 0.23 & 0.13 $\pm$ 0.18\\
HD 183915       & $-$0.39 $\pm$ 0.14 & 0.76 $\pm$ 0.31 & 1.08 $\pm$ 0.25 & 0.83 $\pm$ 0.25 & 0.50 $\pm$ 0.30 & 1.12 $\pm$ 0.21 & 0.75 $\pm$ 0.24 & 0.31 $\pm$ 0.19\\
HD 187308       & $-$0.08 $\pm$ 0.11 & 0.60 $\pm$ 0.28 & 0.89 $\pm$ 0.25 & 0.66 $\pm$ 0.25 & 0.72 $\pm$ 0.26 & 0.57 $\pm$ 0.20 & 0.45 $\pm$ 0.23 & 0.08 $\pm$ 0.17\\
HD 193530       & $-$0.17 $\pm$ 0.14 & 0.38  & 1.05  & 0.95 $\pm$ 0.25 & 0.35  & 0.49 $\pm$ 0.21 & 0.50 $\pm$ 0.25 &$-$0.03 $\pm$ 0.19\\
HD 196445       & $-$0.19 $\pm$ 0.17 & 0.99 $\pm$ 0.33 & 1.23 $\pm$ 0.25 & 1.10 $\pm$ 0.25 & 0.77 $\pm$ 0.31 & 0.98 $\pm$ 0.22 & 0.86 $\pm$ 0.24 & 0.16 $\pm$ 0.19\\
HD 199435       & $-$0.39 $\pm$ 0.12 & 0.91 $\pm$ 0.26 & 1.36  & 0.96 $\pm$ 0.24 & 1.27  & 1.16 $\pm$ 0.20 & 0.97 $\pm$ 0.23 & 0.41 $\pm$ 0.18\\
HD 200995       & $-$0.03 $\pm$ 0.17 & 0.56  & 0.92 $\pm$ 0.25 & 0.52 $\pm$ 0.26 & 0.46 $\pm$ 0.31 & 0.56 $\pm$ 0.20 & 0.49 $\pm$ 0.24 & 0.15 $\pm$ 0.19\\
HD 201657       & $-$0.34 $\pm$ 0.17 & 1.01 $\pm$ 0.30 & 1.41 $\pm$ 0.25 & 1.08 $\pm$ 0.24 & 0.96 $\pm$ 0.25 & 1.27 $\pm$ 0.20 & 1.10 $\pm$ 0.24 & 0.45 $\pm$ 0.17\\

HD 201824       & $-$0.33 $\pm$ 0.17 & 0.98 $\pm$ 0.28 & 1.20 $\pm$ 0.27 & 0.84 $\pm$ 0.24 & 0.78  & 1.25 $\pm$ 0.20 & 1.06 $\pm$ 0.24 & 0.35 $\pm$ 0.17\\
HD 204075       & $+$0.06 $\pm$ 0.17 & 0.86  & 1.59  & 1.02 $\pm$ 0.23 & $...$ & 0.76 $\pm$ 0.21 & 0.38 $\pm$ 0.24 & 0.05 $\pm$ 0.18\\
HD 207277       & $-$0.13 $\pm$ 0.14 & 0.64 $\pm$ 0.34 & 0.87 $\pm$ 0.25 & 0.64 $\pm$ 0.25 & 0.34 $\pm$ 0.31 & 0.81 $\pm$ 0.22 & 0.88 $\pm$ 0.24 & 0.11 $\pm$ 0.19\\
HD 210709       & $-$0.10 $\pm$ 0.14 & 0.42 $\pm$ 0.30 & 0.76 $\pm$ 0.26 & 0.59 $\pm$ 0.24 & 0.27  & 0.64 $\pm$ 0.20 & 0.70 $\pm$ 0.26 & 0.05 $\pm$ 0.17\\
HD 210946       & $-$0.12 $\pm$ 0.13 & 0.61 $\pm$ 0.30 & 0.86 $\pm$ 0.25 & 0.66 $\pm$ 0.24 & 0.62 $\pm$ 0.26 & 0.54 $\pm$ 0.20 & 0.33 $\pm$ 0.24 &$-$0.07 $\pm$ 0.17\\
HD 211173       & $-$0.39 $\pm$ 0.09 & 0.48  & 0.71 $\pm$ 0.25 & 0.52 $\pm$ 0.25 & 0.20 $\pm$ 0.25 & 0.39 $\pm$ 0.21 & 0.20 $\pm$ 0.23 &$-$0.09 $\pm$ 0.17\\
HD 211594       & $-$0.43 $\pm$ 0.14 & 1.21 $\pm$ 0.27 & 1.56 $\pm$ 0.25 & 1.26 $\pm$ 0.24 & 1.05 $\pm$ 0.27 & 1.43 $\pm$ 0.21 & 1.02 $\pm$ 0.24 & 0.57 $\pm$ 0.17\\
HD 211954       & $-$0.51 $\pm$ 0.19 & 0.95 $\pm$ 0.34 & 1.17 $\pm$ 0.26 & 1.06 $\pm$ 0.26 & 0.70 $\pm$ 0.30 & 1.35 $\pm$ 0.22 & 1.47 $\pm$ 0.24 & 0.38 $\pm$ 0.19\\
HD 214579       & $-$0.26 $\pm$ 0.14 & 0.51 $\pm$ 0.32 & 1.03 $\pm$ 0.25 & 0.70 $\pm$ 0.25 & 0.47 $\pm$ 0.30 & 0.72 $\pm$ 0.21 & 0.58 $\pm$ 0.24 & 0.23 $\pm$ 0.19\\
HD 217143       & $-$0.35 $\pm$ 0.17 & 0.62 $\pm$ 0.31 & 0.97 $\pm$ 0.25 & 0.85 $\pm$ 0.25 & 0.56 $\pm$ 0.30 & 0.93 $\pm$ 0.21 & 0.75 $\pm$ 0.24 & 0.16 $\pm$ 0.19\\

HD 217447       & $-$0.17 $\pm$ 0.11 & 0.92 $\pm$ 0.26 & 1.02  & 0.93 $\pm$ 0.24 & 0.80 $\pm$ 0.25 & 0.79 $\pm$ 0.20 & 0.54 $\pm$ 0.22 & 0.08 $\pm$ 0.18\\
HD 219116       & $-$0.61 $\pm$ 0.09 & 0.81  & 1.28  & 0.75 $\pm$ 0.25 & 1.27  & 0.98 $\pm$ 0.20 & 0.75 $\pm$ 0.23 & 0.25 $\pm$ 0.17\\
HD 223586       & $-$0.08 $\pm$ 0.11 & 0.75 $\pm$ 0.28 & 1.08 $\pm$ 0.25 & 0.87 $\pm$ 0.24 & 0.71 $\pm$ 0.26 & 0.84 $\pm$ 0.20 & 0.79 $\pm$ 0.25 & 0.29 $\pm$ 0.17\\
HD 223617       & $-$0.18 $\pm$ 0.13 & 0.58 $\pm$ 0.30 & 1.03 $\pm$ 0.25 & 0.94 $\pm$ 0.25 & 0.69 $\pm$ 0.26 & 0.72 $\pm$ 0.20 & 0.47 $\pm$ 0.24 & 0.13 $\pm$ 0.17\\
HD 252117       & $-$0.14 $\pm$ 0.19 & 0.79  & 1.06 $\pm$ 0.26 & 1.01 $\pm$ 0.26 & 0.79 $\pm$ 0.31 & 0.97 $\pm$ 0.21 & 0.77 $\pm$ 0.24 & 0.19 $\pm$ 0.19\\
HD 273845       & $-$0.15 $\pm$ 0.16 & 0.74  & 1.02 $\pm$ 0.26 & 0.97 $\pm$ 0.25 & 0.67 $\pm$ 0.25 & 1.14 $\pm$ 0.20 & 0.89 $\pm$ 0.23 & 0.35 $\pm$ 0.17\\
HD 288174       & $-$0.05 $\pm$ 0.15 & 0.33  & 0.87  & 0.66 $\pm$ 0.25 & 0.54 $\pm$ 0.26 & 0.74 $\pm$ 0.21 & 0.30 $\pm$ 0.24 & 0.11 $\pm$ 0.17\\
MFU 112         & $-$0.43 $\pm$ 0.15 & 1.18 $\pm$ 0.29 & 1.78  & 1.55 $\pm$ 0.26 & 1.06  & 1.53 $\pm$ 0.21 & 1.13 $\pm$ 0.23 & 0.65 $\pm$ 0.17\\
BD$-18^{\circ}$ 821       & $-$0.27 $\pm$ 0.15 & 0.66 $\pm$ 0.28 & 0.72  & 0.79 $\pm$ 0.24 & 0.58  & 0.82 $\pm$ 0.20 & 0.99 $\pm$ 0.23 & 0.27 $\pm$ 0.18\\
CD$-26^{\circ}$ 7844      & $+$0.02 $\pm$ 0.11 & 0.54 $\pm$ 0.26 & 0.37  & 0.38 $\pm$ 0.25 & 0.37  & 0.29 $\pm$ 0.21 & 0.24 $\pm$ 0.23 & 0.01 $\pm$ 0.18\\

CD$-30^{\circ}$ 9005      & $+$0.05 $\pm$ 0.12 & 0.55 $\pm$ 0.27 & 0.73 $\pm$ 0.25 & 0.74 $\pm$ 0.25 & 0.59 $\pm$ 0.27 & 0.65 $\pm$ 0.21 & 0.49 $\pm$ 0.24 & 0.11 $\pm$ 0.17\\
CD$-34^{\circ}$ 6139      & $-$0.07 $\pm$ 0.13 & 0.51 $\pm$ 0.30 & 0.87  & 1.17 $\pm$ 0.25 & 0.89 $\pm$ 0.28 & 0.60 $\pm$ 0.20 & 0.38 $\pm$ 0.23 & 0.06 $\pm$ 0.17\\
CD$-34^{\circ}$ 7430      & $+$0.01 $\pm$ 0.14 & 0.66  & 0.56  & 0.45 $\pm$ 0.24 & 0.33  & 0.50 $\pm$ 0.21 & 0.30 $\pm$ 0.24 &$-$0.01 $\pm$ 0.17\\
CD$-46^{\circ}$ 3977      & $-$0.10 $\pm$ 0.15 & 0.65  & 0.91  & 0.76 $\pm$ 0.25 & 0.69 $\pm$ 0.29 & 0.64 $\pm$ 0.20 & 0.40 $\pm$ 0.24 & 0.18 $\pm$ 0.17\\
HD 18182        & $-$0.17 $\pm$ 0.10 & 0.70  & 0.38  & 0.80 $\pm$ 0.27 & 0.70  & 0.37 $\pm$ 0.21 & 0.28 $\pm$ 0.24 & 0.01 $\pm$ 0.17\\
HD 18361        & $+$0.01 $\pm$ 0.15 & 0.30  & 0.71  & 0.47 $\pm$ 0.24 & 0.69  & 0.48 $\pm$ 0.22 & 0.16 $\pm$ 0.24 & 0.02 $\pm$ 0.17\\
HD 21682        & $-$0.48 $\pm$ 0.12 & 0.37  & $...$ & 0.56  & $...$ & 0.81 $\pm$ 0.20 & 0.68 $\pm$ 0.23 & 0.40 $\pm$ 0.18\\
HD 26886        & $-$0.30 $\pm$ 0.10 & 0.40  & $...$ & 0.41  & 0.85  & 0.62 $\pm$ 0.20 & 0.40 $\pm$ 0.23 & 0.07 $\pm$ 0.18\\
HD 31812        & $-$0.07 $\pm$ 0.11 & 0.69  & 0.66  & 0.62 $\pm$ 0.23 & 0.74 $\pm$ 0.25 & 0.51 $\pm$ 0.20 & 0.28 $\pm$ 0.23 &$-$0.09 $\pm$ 0.18\\
HD 33709        & $-$0.20 $\pm$ 0.14 & 0.35  & $...$ & 0.30  & 0.88  & 0.40 $\pm$ 0.20 & 0.17 $\pm$ 0.22 & 0.07 $\pm$ 0.18\\

HD 39778        & $-$0.12 $\pm$ 0.12 & 0.75  & 0.48  & 0.80 $\pm$ 0.23 & 0.96  & 0.85 $\pm$ 0.20 & 0.59 $\pm$ 0.22 & 0.08 $\pm$ 0.18\\
HD 41701        & $+$0.02 $\pm$ 0.13 & 0.36  & $...$ & 0.58 $\pm$ 0.26 & 0.30  & 0.32 $\pm$ 0.21 & 0.22 $\pm$ 0.23 & 0.05 $\pm$ 0.18\\
HD 45483        & $-$0.14 $\pm$ 0.12 & 0.61  & 0.82  & 0.77 $\pm$ 0.25 & 0.49  & 0.54 $\pm$ 0.21 & 0.33 $\pm$ 0.24 & 0.01 $\pm$ 0.17\\
HD 48814        & $-$0.07 $\pm$ 0.11 & 0.38  & 0.36  & 0.35 $\pm$ 0.24 & 0.20  & 0.24 $\pm$ 0.21 & 0.13 $\pm$ 0.24 &$-$0.04 $\pm$ 0.17\\
HD 49017        & $+$0.02 $\pm$ 0.11 & 0.18  & 0.67  & 0.24 $\pm$ 0.24 & 0.29  & 0.32 $\pm$ 0.21 & 0.27 $\pm$ 0.23 &$-$0.04 $\pm$ 0.18\\
HD 49661        & $-$0.13 $\pm$ 0.10 & 0.34  & 0.63  & 0.18 $\pm$ 0.24 & 0.42  & 0.20 $\pm$ 0.20 & 0.13 $\pm$ 0.22 & $...$\\
HD 49778        & $-$0.22 $\pm$ 0.12 &$-$0.18  & 0.82  & 0.35 $\pm$ 0.23 & 0.53  & 0.52 $\pm$ 0.21 & 0.31 $\pm$ 0.23 & 0.29 $\pm$ 0.18\\
HD 50075        & $-$0.16 $\pm$ 0.11 & 0.44  & 0.93 $\pm$ 0.25 & 0.58 $\pm$ 0.24 & 0.85 $\pm$ 0.25 & 0.81 $\pm$ 0.20 & 0.67 $\pm$ 0.23 & 0.24 $\pm$ 0.17\\
HD 50843        & $-$0.31 $\pm$ 0.13 & 0.45  & 0.60  & 0.31 $\pm$ 0.24 & 0.26 $\pm$ 0.26 & 0.50 $\pm$ 0.20 & 0.40 $\pm$ 0.23 & 0.03 $\pm$ 0.17\\
HD 53199        & $-$0.23 $\pm$ 0.13 & 0.75  & 0.95  & 0.55 $\pm$ 0.23 & 0.75 $\pm$ 0.23 & 0.77 $\pm$ 0.20 & 0.69 $\pm$ 0.24 & 0.12 $\pm$ 0.18\\

HD 58121        & $-$0.01 $\pm$ 0.13 & 0.38 $\pm$ 0.31 & 0.61 $\pm$ 0.25 & 0.34 $\pm$ 0.25 & 0.01 $\pm$ 0.30 & 0.26 $\pm$ 0.21 & 0.16 $\pm$ 0.25 &$-$0.21 $\pm$ 0.19\\
HD 88495        & $-$0.11 $\pm$ 0.10 & 0.96 $\pm$ 0.29 & 1.03 $\pm$ 0.25 & 0.73 $\pm$ 0.24 & 0.38  & 0.55 $\pm$ 0.21 & 0.44 $\pm$ 0.28 &$-$0.11 $\pm$ 0.17\\
HD 90167        & $-$0.04 $\pm$ 0.11 & 0.48 $\pm$ 0.27 & 0.77  & 0.44 $\pm$ 0.23 & 0.47 $\pm$ 0.24 & 0.29 $\pm$ 0.20 & 0.28 $\pm$ 0.23 & 0.10 $\pm$ 0.18\\
HD 95193        & $+$0.04 $\pm$ 0.12 & 0.49 $\pm$ 0.26 & 0.86 $\pm$ 0.26 & 0.54 $\pm$ 0.23 & 0.37 $\pm$ 0.24 & 0.53 $\pm$ 0.20 & 0.26 $\pm$ 0.23 & 0.11 $\pm$ 0.18\\
HD 107270       & $+$0.05 $\pm$ 0.17 & 0.13  & 1.17  & 1.20 $\pm$ 0.23 & 0.59  & 0.41 $\pm$ 0.21 & 0.44 $\pm$ 0.28 & 0.15 $\pm$ 0.18\\
HD 109061       & $-$0.56 $\pm$ 0.09 & 0.54  & 0.86  & 0.55 $\pm$ 0.24 & 0.32 $\pm$ 0.25 & 0.64 $\pm$ 0.20 & 0.64 $\pm$ 0.24 & 0.31 $\pm$ 0.17\\
HD 113195       & $-$0.15 $\pm$ 0.12 & 0.59 $\pm$ 0.28 & 0.79 $\pm$ 0.26 & 0.59 $\pm$ 0.24 & 0.54 $\pm$ 0.26 & 0.41 $\pm$ 0.21 & 0.45 $\pm$ 0.24 & 0.10 $\pm$ 0.17\\
HD 115277       & $-$0.03 $\pm$ 0.15 & 0.53 $\pm$ 0.28 & 0.82 $\pm$ 0.25 & 0.48 $\pm$ 0.24 & 0.27 $\pm$ 0.25 & 0.33 $\pm$ 0.21 & 0.35 $\pm$ 0.24 &$-$0.02 $\pm$ 0.17\\
HD 119650       & $-$0.10 $\pm$ 0.13 & 0.31 $\pm$ 0.32 & 0.45  & 0.33 $\pm$ 0.25 &$-$0.05 $\pm$ 0.30 & 0.18 $\pm$ 0.21 & 0.18 $\pm$ 0.24 &$-$0.14 $\pm$ 0.19\\
HD 134698       & $-$0.52 $\pm$ 0.12 & 0.39  & 0.97 $\pm$ 0.25 & 0.51 $\pm$ 0.26 & 0.36  & 0.46 $\pm$ 0.21 & 0.25 $\pm$ 0.24 & 0.08 $\pm$ 0.19\\

HD 139266       & $-$0.27 $\pm$ 0.18 & 0.66 $\pm$ 0.32 & 0.94 $\pm$ 0.25 & 0.62 $\pm$ 0.25 & 0.43 $\pm$ 0.30 & 0.75 $\pm$ 0.21 & 0.71 $\pm$ 0.25 & 0.16 $\pm$ 0.19\\
HD 139409       & $-$0.51 $\pm$ 0.13 & 0.49 $\pm$ 0.27 & 0.60  & 0.58 $\pm$ 0.24 & 0.47 $\pm$ 0.26 & 0.43 $\pm$ 0.20 & 0.40 $\pm$ 0.24 & 0.03 $\pm$ 0.17\\
HD 169106       & $+$0.01 $\pm$ 0.12 & 0.40  & $...$ & 0.42 $\pm$ 0.24 & 0.29 $\pm$ 0.26 & 0.33 $\pm$ 0.20 & 0.38 $\pm$ 0.24 & 0.10 $\pm$ 0.17\\
HD 184001       & $-$0.21 $\pm$ 0.14 & 0.62 $\pm$ 0.28 & 0.97  & 0.59 $\pm$ 0.23 & 0.68 $\pm$ 0.24 & 0.59 $\pm$ 0.20 & 0.40 $\pm$ 0.23 & 0.09 $\pm$ 0.18\\
HD 204886       & $+$0.04 $\pm$ 0.15 & 0.42 $\pm$ 0.34 & 0.89 $\pm$ 0.25 & 0.63 $\pm$ 0.26 & 0.42 $\pm$ 0.31 & 0.64 $\pm$ 0.21 & 0.46 $\pm$ 0.24 & 0.05 $\pm$ 0.19\\
HD 213084       & $-$0.09 $\pm$ 0.15 & 0.66 $\pm$ 0.27 & 0.89  & 0.78 $\pm$ 0.23 & 0.75 $\pm$ 0.24 & 0.88 $\pm$ 0.20 & 0.67 $\pm$ 0.22 & 0.18 $\pm$ 0.18\\
HD 223938       & $-$0.42 $\pm$ 0.11 & 0.63  & $...$ & 0.62 $\pm$ 0.23 & 0.77 $\pm$ 0.23 & 0.78 $\pm$ 0.20 & 0.64 $\pm$ 0.23 & 0.08 $\pm$ 0.18\\
MFU 214         & $+$0.00 $\pm$ 0.12 & 0.35 $\pm$ 0.27 & 0.37  & 0.34 $\pm$ 0.24 & 0.05 $\pm$ 0.26 & 0.22 $\pm$ 0.21 & 0.11 $\pm$ 0.24 &$-$0.15 $\pm$ 0.17\\
MFU 229         & $-$0.01 $\pm$ 0.11 & 0.63  & 0.67  & 0.49 $\pm$ 0.24 & 0.44 $\pm$ 0.26 & 0.49 $\pm$ 0.20 & 0.35 $\pm$ 0.24 & 0.13 $\pm$ 0.17\\
HD 12392        & $-$0.08 $\pm$ 0.18 & 0.73 $\pm$ 0.27 & 1.13  & 0.96 $\pm$ 0.24 & 0.74 $\pm$ 0.26 & 1.27 $\pm$ 0.21 & 1.19 $\pm$ 0.25 & 0.41 $\pm$ 0.17\\

HD 17067        & $-$0.61 $\pm$ 0.21 & 0.95  & 1.31 $\pm$ 0.26 & 0.76 $\pm$ 0.25 & 0.67 $\pm$ 0.31 & 0.93 $\pm$ 0.21 & 0.73 $\pm$ 0.24 & 0.37 $\pm$ 0.19\\
HD 90127        & $-$0.40 $\pm$ 0.10 & 0.99  & 1.17 $\pm$ 0.25 & 0.87 $\pm$ 0.26 & 0.60 $\pm$ 0.30 & 0.80 $\pm$ 0.21 & 0.46 $\pm$ 0.24 & 0.25 $\pm$ 0.19\\
HD 102762       & $-$0.17 $\pm$ 0.20 & 0.74 $\pm$ 0.32 & 1.14 $\pm$ 0.25 & 0.93 $\pm$ 0.25 & 0.57 $\pm$ 0.30 & 1.11 $\pm$ 0.21 & 0.90 $\pm$ 0.24 & 0.31 $\pm$ 0.19\\
HD 114678       & $-$0.50 $\pm$ 0.13 & 1.11 $\pm$ 0.26 & $...$ & 1.09 $\pm$ 0.23 & 0.79 $\pm$ 0.25 & 1.21 $\pm$ 0.20 & 0.97 $\pm$ 0.22 & 0.42 $\pm$ 0.18\\
HD 180622       & $+$0.03 $\pm$ 0.12 & 0.55 $\pm$ 0.32 & 0.72 $\pm$ 0.26 & 0.52 $\pm$ 0.26 & 0.10 $\pm$ 0.31 & 0.33 $\pm$ 0.22 & 0.28 $\pm$ 0.24 &$-$0.10 $\pm$ 0.19\\
HD 200063       & $-$0.34 $\pm$ 0.20 & 0.84  & 1.02 $\pm$ 0.25 & 0.61 $\pm$ 0.25 & 0.40 $\pm$ 0.31 & 0.81 $\pm$ 0.21 & 0.58 $\pm$ 0.24 & 0.27 $\pm$ 0.19\\
HD 210030       & $-$0.03 $\pm$ 0.11 & 0.44 $\pm$ 0.28 & 0.47  & 0.46 $\pm$ 0.25 &$-$0.03 $\pm$ 0.25 & 0.22 $\pm$ 0.21 & 0.14 $\pm$ 0.24 &$-$0.22 $\pm$ 0.17\\
HD 214889       & $-$0.17 $\pm$ 0.12 & 0.62  & 0.65  & 0.62 $\pm$ 0.24 & 0.43 $\pm$ 0.27 & 0.59 $\pm$ 0.20 & 0.37 $\pm$ 0.23 & 0.06 $\pm$ 0.17\\
HD 215555       & $-$0.08 $\pm$ 0.12 & 0.99  & 0.98  & 0.98 $\pm$ 0.23 & 0.87 $\pm$ 0.24 & 0.72 $\pm$ 0.20 & 0.47 $\pm$ 0.23 &$-$0.05 $\pm$ 0.18\\
HD 216809       & $-$0.04 $\pm$ 0.14 & 0.79 $\pm$ 0.32 & 0.92 $\pm$ 0.25 & 0.61 $\pm$ 0.25 & 0.19 $\pm$ 0.30 & 0.16 $\pm$ 0.21 & 0.24 $\pm$ 0.25 &$-$0.29 $\pm$ 0.19\\

HD 221879       & $-$0.10 $\pm$ 0.19 & 0.74  & 1.07 $\pm$ 0.25 & 0.60 $\pm$ 0.26 & 0.30 $\pm$ 0.31 & 0.38 $\pm$ 0.20 & 0.26 $\pm$ 0.24 &$-$0.12 $\pm$ 0.19\\
HD 749          & $-$0.29 $\pm$ 0.15 & 0.72 $\pm$ 0.29 & 1.21 $\pm$ 0.25 & 0.97 $\pm$ 0.24 & 1.05 $\pm$ 0.26 & 0.98 $\pm$ 0.20 & 0.88 $\pm$ 0.24 & 0.28 $\pm$ 0.17\\
HD 88927        & $+$0.02 $\pm$ 0.13 & 0.57 $\pm$ 0.34 & 0.48 $\pm$ 0.26 & 0.32 $\pm$ 0.26 & 0.03 $\pm$ 0.30 & 0.20 $\pm$ 0.21 & 0.25 $\pm$ 0.24 &$-$0.17 $\pm$ 0.19\\
BD$+09^{\circ}$ 2384      & $-$0.98 $\pm$ 0.10 & $...$ & $...$ & 0.75  & $...$ & 0.70 $\pm$ 0.19 & 0.77 $\pm$ 0.24 & 0.48 $\pm$ 0.17\\
HD 89638        & $-$0.19 $\pm$ 0.11 & 0.64  & 0.29  & 0.63 $\pm$ 0.25 & 0.75  & 0.57 $\pm$ 0.20 & 0.36 $\pm$ 0.23 & 0.03 $\pm$ 0.17\\
HD 187762       & $-$0.30 $\pm$ 0.11 & 0.44  & $...$ & 0.36 $\pm$ 0.24 & 0.38  & 0.48 $\pm$ 0.20 & 0.51 $\pm$ 0.24 & 0.33 $\pm$ 0.17\\
NGC5 822-201    & $-$0.11 $\pm$ 0.10 & 0.83  & 1.03  & 0.74  & 0.97  & 0.72 $\pm$ 0.20 & 0.47 $\pm$ 0.23 & 0.17 $\pm$ 0.18\\
NGC 5822-2      & $-$0.15 $\pm$ 0.09 & 0.75  & 0.43  & 0.90  & $...$ & 0.67 $\pm$ 0.20 & 0.34 $\pm$ 0.23 & 0.31 $\pm$ 0.18\\
HD 10613        & $-$0.92 $\pm$ 0.12 & 1.10  & 1.36  & 0.94  & 1.38  & 1.28 $\pm$ 0.20 & 1.31 $\pm$ 0.24 & 0.74 $\pm$ 0.17\\
CD$-25^{\circ}$ 6606      & $+$0.12 $\pm$ 0.14 & 0.58  & $...$ & $...$ & $...$ & 0.45 $\pm$ 0.21 & 0.21 $\pm$ 0.24 & 0.13 $\pm$ 0.18\\

HD 46040        & $+$0.11 $\pm$ 0.13 & 0.81 $\pm$ 0.27 & 1.22 $\pm$ 0.25 & 1.03 $\pm$ 0.25 & 0.81 $\pm$ 0.26 & 1.02 $\pm$ 0.21 & 0.77 $\pm$ 0.24 & 0.21 $\pm$ 0.17\\
HD 49841        & $+$0.21 $\pm$ 0.13 & 0.46  & 0.74  & 0.77 $\pm$ 0.24 & 0.47  & 0.61 $\pm$ 0.21 & 0.45 $\pm$ 0.23 & 0.07 $\pm$ 0.18\\
HD 82765        & $+$0.19 $\pm$ 0.10 & 0.34  & $...$ & 0.79 $\pm$ 0.26 & $...$ & 0.25 $\pm$ 0.21 & 0.31 $\pm$ 0.25 & 0.01 $\pm$ 0.18\\
HD 84734        & $+$0.20 $\pm$ 0.12 & 0.45  & 0.78  & 0.90 $\pm$ 0.27 & 0.58 $\pm$ 0.24 & 0.54 $\pm$ 0.21 & 0.21 $\pm$ 0.23 & 0.17 $\pm$ 0.18\\
HD 85205        & $+$0.23 $\pm$ 0.16 & 0.62  & 0.46  & $...$ & $...$ & 0.45 $\pm$ 0.21 & 0.19 $\pm$ 0.24 &$-$0.03 $\pm$ 0.18\\
HD 101079       & $+$0.10 $\pm$ 0.12 & 0.41  & 0.40  & 0.44 $\pm$ 0.24 & 0.33  & 0.48 $\pm$ 0.21 & 0.17 $\pm$ 0.23 &$-$0.10 $\pm$ 0.18\\
HD 130386       & $+$0.16 $\pm$ 0.13 & 0.53  & 0.79 $\pm$ 0.26 & 0.79 $\pm$ 0.25 & 0.24  & 0.36 $\pm$ 0.22 & 0.27 $\pm$ 0.25 &$-$0.08 $\pm$ 0.17\\
HD 139660       & $+$0.26 $\pm$ 0.14 & 0.39  & 0.54  & 0.69 $\pm$ 0.25 & 0.17  & 0.31 $\pm$ 0.22 & 0.41 $\pm$ 0.24 &$-$0.08 $\pm$ 0.18\\
HD 198590       & $+$0.18 $\pm$ 0.14 & 0.43  & 0.66  & 0.46  & $...$ & 0.22 $\pm$ 0.22 & 0.09 $\pm$ 0.23 & 0.03 $\pm$ 0.18\\
HD 212209       & $+$0.30 $\pm$ 0.13 & 0.32  & 0.40  & 0.16 $\pm$ 0.25 &$-$0.08  & 0.15 $\pm$ 0.22 & 0.19 $\pm$ 0.25 &$-$0.23 $\pm$ 0.17\\
\end{longtable}

\clearpage
\twocolumn

\bsp	
\label{lastpage}
\end{document}